\documentclass[prb,aps,epsf,twocolumn,showpacs,10pt]{revtex4-1}
\usepackage{graphicx,amsfonts,times,bm,amsmath,verbatim,color,array}

\renewcommand{\>}{\rangle}
\renewcommand{\(}{\left(}
\renewcommand{\)}{\right)}
\renewcommand{\[}{\left[}
\renewcommand{\]}{\right]}

\begin{document}
\title{Nernst and magneto-thermal conductivity in a lattice model of Weyl fermions}
\author{Girish Sharma$^1$}
\author{Pallab Goswami$^2$}
\author{Sumanta Tewari$^1$}
\affiliation{$^1$Department of Physics and Astronomy, Clemson University, Clemson, SC 29634\\
$^2$Condensed Matter Theory Center, University of Maryland, College Park, Maryland 20742}
\begin{abstract}
Weyl semimetals (WSM) are topologically protected three dimensional materials whose low energy excitations are linearly dispersing massless Dirac fermions, possessing a non-trivial Berry curvature. Using semi-classical Boltzmann dynamics in the relaxation time approximation for a lattice model of time reversal (TR) symmetry broken WSM, we compute both magnetic field dependent and anomalous contributions to the Nernst coefficient. In addition to the magnetic field dependent Nernst response, which is present in both Dirac and Weyl semimetals, we show that, contrary to previous reports, the TR-broken WSM also has an  anomalous Nernst response due to a non-vanishing Berry curvature. 
We also compute the thermal conductivities of a WSM in the Nernst (${\nabla T} \perp \mathbf{B}$) and the longitudinal (${\nabla T} \parallel \mathbf{B}$)  set-up and confirm from our lattice model that in the parallel set-up, the Wiedemann-Franz law is violated between the longitudinal thermal and electrical conductivities due to chiral anomaly.  
\end{abstract}
\maketitle
\section{Introduction}
After the theoretical prediction of topological insulators, and their subsequent experimental realization, the field of topological condensed matter has grown manifold~\cite{Bernevig:2006, Konig:2007, LFu:2007, Kane:2005, Zhang:2009, Hasan:2010, Moore:2010, Qi:2011, Roy:2009}. The topological order manifested in these systems is not associated with spontaneous breaking of a symmetry, but rather can be described by a topological invariant which is insensitive to a smooth deformation of the Hamiltonian. Usually the robust topological protection is associated with a non-zero spectral gap in the bulk of the system, and the presence of protected zero energy surface states is regarded as the hallmark of a non-trivial topological phase of matter. However, recently it has been proposed that systems in three spatial dimensions, in the presence of broken time-reversal (TR) and/or space-inversion (SI) symmetry, can also be topologically protected even without a bulk energy gap~\cite{Volovik:2001, Wan:2011, Yang:2011, Burkov:2011, Burkov2:2011, Xu:2011, Zyuzin:2012, Zyuzin2:2012, Meng:2012, Gong:2011, Sau:2012, Hosur:2013}. These are Weyl semimetals (WSM) - the nomenclature based on the Dirac/Weyl equation which is used to describe their low energy excitations~\cite{Peskin}. 

A number of recent experiments have claimed to be able to observe the Weyl semimetal phase in an inversion asymmetric compound TaAs~\cite{Su:2015, Huang:2015, Lv:2015}, and also in a 3D double gyroid photonic crystal~\cite{Llu:2015}, without breaking TR. Another route which can result in the experimental verification of the novel WSM phase is to first realize a 3D Dirac semimetal and then break time reversal symmetry by applying a magnetic field, which will split a Dirac cone into a pair of Weyl nodes. Na$_3$Bi and Cd$_3$As$_2$ were recently proposed to be Dirac semimetals~\cite{Wang:2012, Wang:2013}, and also have been confirmed experimentally by a series of experiments~\cite{Liu:2014, SyXu:2013, Neupane:2014, ZKLiu:2014, Borisenko:2014, Jeon:2014, TLiang:2014, Xiong1:2015, Xiong:2015}. In Bi$_{1-x}$Sb$_x$ for $x\sim 3-4\%$ also the Dirac semimetal phase has been predicted~\cite{Fu:2007, Teo:2008, Guo:2011}, and experimental signatures of realizing a WSM phase by breaking TR have been reported~\cite{Kim:2013}.

A simple WSM with broken time reversal symmetry can be desribed by a pair of linearly dispersing massless Dirac fermions governed by the Hamiltonian: $H_{\pm}(\mathbf{k}) = \pm \hbar v_F\boldsymbol\sigma\cdot(\mathbf{k}-\mathbf{K}_{\pm})$, where $\boldsymbol\sigma$ is the vector of Pauli spin matrices defined in the space of two non-degenerate energy bands, $v_F$ is the Fermi velocity, and $\mathbf{K}_{\pm}$ are the two band touching points separated from each other in momentum space by $\mathbf{k}_0=\mathbf{K}_+-\mathbf{K}_-$. It is essential that $\mathbf{k}_0$ is non-zero to ensure that the system breaks TR and is topologically non-trivial, in which case $H_+$ and $H_-$ describe two Weyl fermions of opposite chirality. The two band touching Weyl points act as a source and a sink (monopole and anti-monopole) of Berry curvature, which acts as a fictitious magnetic field on the electron wave-function in momentum space~\cite{Niu:2006}. For $\mathbf{k}_0=0$, the two Weyl points collapse onto each other giving rise to a topologically trivial (i.e with a zero Berry curvature flux) massless degenerate Dirac fermion. The topological nature of a WSM leads to a host of interesting physics, for example Berry curvature induced anomalous transport, namely charge and thermal Hall conductivities~\cite{Burkov:2014, Goswami:2013, Fiete:2014, Kim:2014, Kim2:2014, Kim3:2014} and open Fermi arcs on surfaces~\cite{Wan:2011, Hosur:2012, Okugawa:2014, Haldane:2014, Potter:2014, Imura:2011, Lu:2015, Delplace:2012}. Anomalous transport phenomena, however, have been already known to exist in a variety of systems which possess a non-trivial distribution of the Berry curvature flux~\cite{Haldane:2004, Niu:2010}. In a WSM, more interestingly, each Weyl node is chiral, with the chirality quantum number protected by a quantized flux of the Berry curvature, also known as Chern flux, which results in another peculiar phenomenon known as chiral anomaly (or Adler-Bell-Jackiw anomaly)~\cite{Volovik:2001, Zyuzin:2012, Nielsen:1983, Aji:2012}. The chiral anomaly concerns with the nonconservation of chiral charge i.e. an imbalance of charge between two distinct species of chiral fermions in the presence of non-orthogonal applied electric and magnetic fields. Several transport signatures have been proposed to test chiral anomaly such as negative longitudinal magenoresistance~\cite{Hosur:2013, Zyuzin:2012, Son:2013, Burkov:2015} and chiral magnetic effect\cite{Zyuzin:2012, Goswami:2014, Franz:2013}  of which the former has been recently claimed to be observed in experiments~\cite{Huang:2015, Xiong:2015, Li:2014, Kim:2013}.
\begin{figure}[h]
\includegraphics[scale=.3]{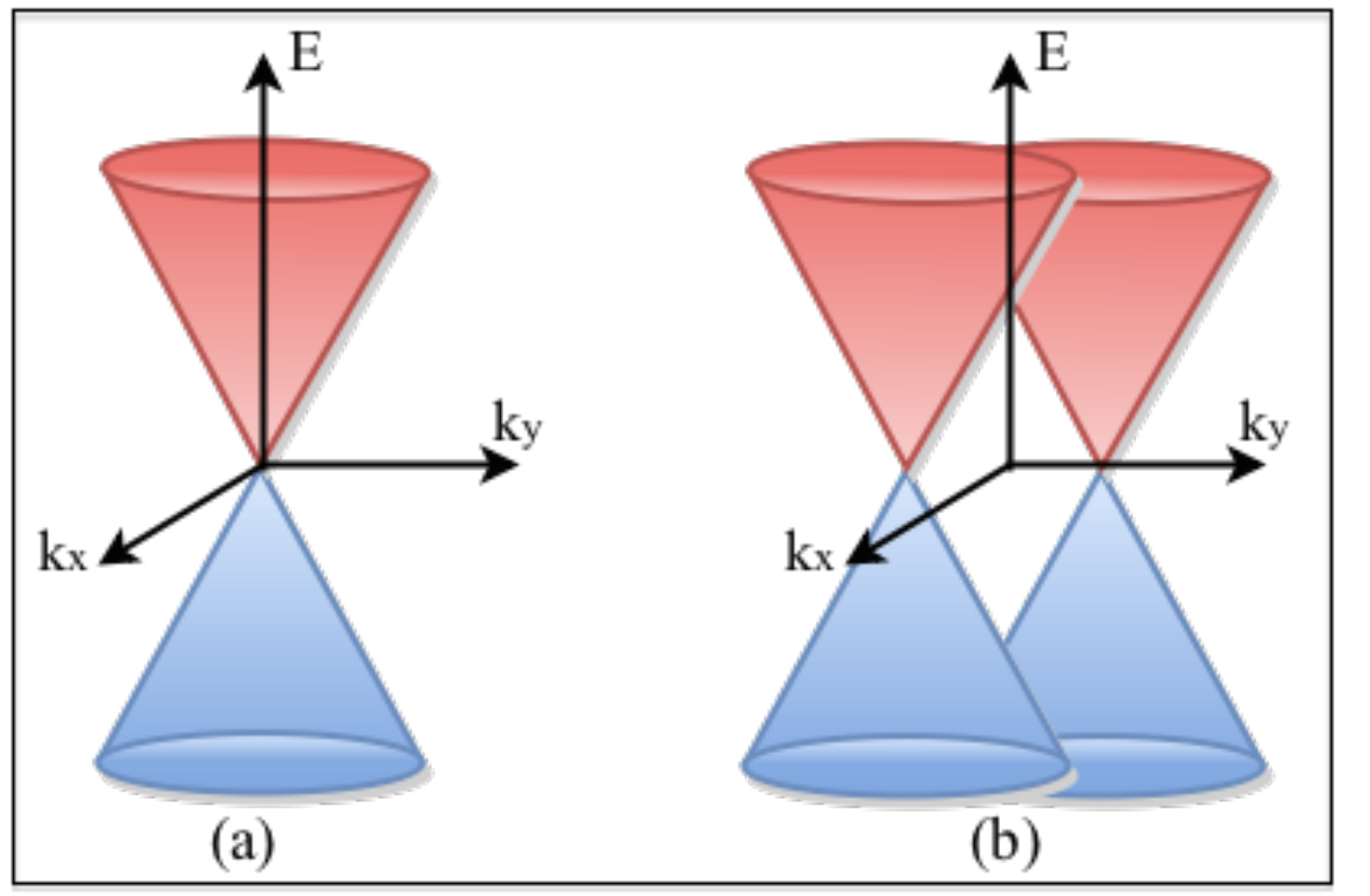}\\
\includegraphics[scale=.24]{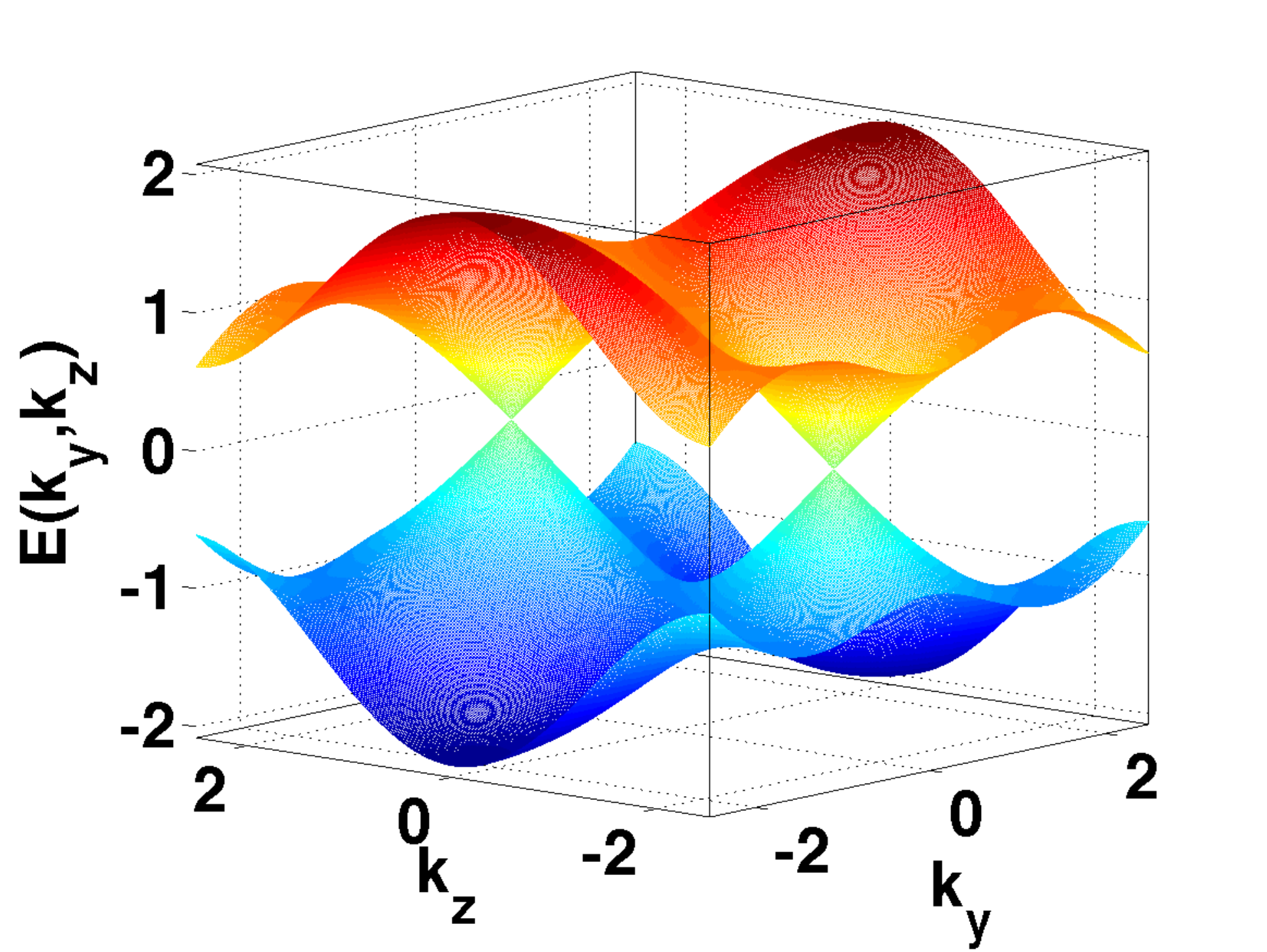}
\caption{\textit{Top:} Linearized band dispersion for Dirac and Weyl semimetals (${k}_z$ is suppressed). (a) A doubly degenerate Dirac semimetal (b) Transition from a Dirac semimetal to a Weyl semimetal (represented by a pair of Dirac cones separated by a finite momentum) by breaking of time-reversal symmetry. \textit{Bottom:} Energy band spectrum for the lattice model of Weyl fermions ($k_x$ suppressed) described in Eq.~\ref{lattice_eqn}. The two band touching points occur at $\mathbf{K_{\pm}}=(0,0,\pm\pi/2)$.}
\label{weyl_2schematic}
\end{figure}

A Dirac node can be split into two Weyl nodes by breaking either the TR symmetry or SI symmetry. Figure~\ref{weyl_2schematic} shows a linearly dispersing Dirac node split into a pair of Weyl nodes when TR symmetry is broken, and also shows the energy-band spectrum of a lattice model of Weyl fermions obtained by diagonalizing the Hamiltonian in Eq.~\ref{lattice_eqn}. The simple model of Weyl semimetal described in the previous paragraph by $H_{\pm}(\mathbf{k})$ breaks TR, however it is also possible to realize a Weyl system when TR is intact but inversion symmetry is broken~\cite{Burkov2:2011, Halasz:2012, Goswami2:2013}. This implies that the system must host more than one flavor of pairs of Weyl fermions for the vector sum of $\mathbf{k}_0$ to vanish. In the SI broken Weyl semimetal, because of TR symmetry there is no Berry curvature induced anomalous charge or thermal Hall effect in the absence of an external magnetic field. However, in the TR broken WSM, because of a finite Berry curvature flux through any plane intermediate between the Weyl nodes in the momentum space, the anomalous charge and thermal Hall conductivities are non-zero~\cite{Goswami:2013, Fiete:2014, Kim:2014, Burkov:2011}.

In this paper, we work with a TR broken phase of WSM and consider its Nernst response. Experimentally, the Nernst effect measures the transverse electrical response to a longitudinal thermal gradient in the presence of a perpendicular magnetic field. The Nernst effect has been used as an important experimental probe in a number of physical systems such as high temperature cuprate superconductors~\cite{Xu:2000, Hagen:1990, Ri:1994, Wang:2006}, and charge density waves~\cite{Bel:2003, Behnia:2009}. Since the TR-broken WSM has a non-zero anomalous Hall response (i.e., non-zero Hall efffect induced by the Berry curvature even in the absence of a magnetic field) it is expected that the anomalous Nernst response will also be non-zero. This is because, the anomalous Hall conductivity $\sigma_{xy}$ and the anomalous Peltier coefficient $\alpha_{xy}$, which measures the transverse electrical currrent in response to a longitudinal temperature gradient, are related by the celebrated Mott relation, $\alpha_{ab} = -\frac{\pi^2}{3}\frac{k_B^2 T}{e}\frac{\partial\sigma_{ab}}{\partial\mu}$. In turn, a non-zero $\alpha_{xy}$ implies an anomalous (zero field) Nernst coefficient given by $\alpha_{xy}/\sigma_{xx}$. In recent work~\cite{Fiete:2014}, however, based on a linearized model of a TR-broken Weyl semimetal, the anomalous Nernst response has been argued to be zero, because a linearized Weyl Hamiltonian with unbounded (or very high) ultraviolet cut-off of the Dirac spectrum produces $\partial \sigma_{xy}/\partial \mu = 0$. Here we show, from a lattice model of a WSM (with the lattice regularization providing a physical ultra-violet cut-off to the low energy Dirac spectrum) that the anomalous Peltier coefficient, and in turn the anomalous Nernst coefficient is finite and measurable in a physical time reversal breaking Weyl semimetal such as Bi$_{1-x}$Sb$_x$. In the main part of the paper, we use the semi-classical Boltzmann equations in the relaxation time approximation in the presence of a non-zero magnetic field and a Berry curvature, and derive the thermoelectric  and charge conductivity tensors (both longitudinal and Hall) which we use to calculate both the conventional (i.e., magnetic field dependent) and topological (i.e. zero field) Nernst coefficients.

Additionally, we also investigate the thermal conductivity of a WSM based on the Boltzmann equation approach. Unlike earlier works which were based on a linearized WSM model,~\cite{Kim:2014, Fiete:2014} we employ a lattice Bloch Hamiltonian. With the Nernst experimental setup i.e. the temperature gradient $\nabla T$ applied perpendicular to the magnetic field $\mathbf{B}$, we find that the transverse magneto-thermal conductivity obeys the Wiedemann-Franz law~\cite{Ziman} (i.e., the ratio of the thermal and electrical conductivity is the Lorenz number $L_0$, both with and without the external magnetic field). In the parallel setup (${\nabla T} \parallel \mathbf{B}$), however, there is an additional $B^2$ dependence of the Lorenz number, thus violating the standard Wiedemann-Franz law for quasiparticles in a Landau Fermi liquid, arising from the chiral anomaly. Our results confirm that the violation of the Wiedemann-Franz law between the longitudinal magneto thermal and electrical conductivities~\cite{Kim2:2014} persist in the physically more transparent lattice model and is not an artifact of the linearized low energy model.

This paper is organized as follows: in Section II, we discuss the Boltzmann semi-classical approach to calculate the Nernst response in a Weyl semimetal. We derive expressions for both longitudinal and transverse charge ($\sigma_{ab}$) and Peltier ($\alpha_{ab}$) conductivity tensors, taking into account perturbative electric and magnetic fields, and a finite temperature gradient, for a Hamiltonian with a non-vanishing Berry curvature. Though the approach is general and can be applied for various configurations, we will compute our expressions relevant for the Nernst experimental setup. Section III concerns with the Nernst response in a linearly dispersing model of Dirac and Weyl fermions. We compute the magnetic field dependent Nernst response for a single Dirac node, and also for a pair of Weyl nodes which have a non-vanishing flux of the Berry curvature. Additionally Weyl fermions also exhibit an anomalous Nernst response even at zero magnetic field when one imposes a physical ultraviolet cut-off on the energy spectrum, and we show that this imposition gives a non-zero Peltier coefficient $\alpha_{xy}$. Section IV concerns with the lattice WSM model and its Nernst response. In Section V, the magneto-thermal conductivity is analyzed, and the Wiedemann-Franz law is studied for orthogonal ($\nabla T\perp\mathbf{B}$) and parallel ($\nabla T\parallel\mathbf{B}$) setups. We conclude in Section VI. 

\section{Boltzmann formalism for Nernst response in a lattice Weyl semimetal}
Nernst effect measures the transverse electrical response to a longitudinal thermal gradient in the presence of a finite magnetic field and absence of a charge current i.e. $E_y=-\vartheta$ $dT/dx$, where $\vartheta$ is defined to be the Nernst coefficient and $-dT/dx$ is the temperature gradient applied along the $x$ axis. The use of three conductivity tensors, $\hat{\sigma}$, $\hat{\alpha}$, and $\hat{l}$ suffices to relate the charge current $\mathbf{J}$ and thermal current $\mathbf{Q}$ to an applied electric field and temperature gradient. We reserve the symbol $\hat{\kappa}$ for thermal conductivity tensor which will be the focus of Section V. We can write the following linear response equation
\begin{equation}
\left( \begin{array}{c}
\mathbf{J} \\
\mathbf{Q}  \\
\end{array} \right) =\left( \begin{array}{cc}
\hat{\sigma} & \hat{\alpha} \\
\hat{\bar{\alpha}} & \hat{l}  \\
\end{array} \right) \left( \begin{array}{c}
\mathbf{E} \\
-\mathbf{\nabla} T  \\
\end{array} \right)
\label{thermal}
\end{equation}
The tensors $\hat{\bar{\alpha}}$ and $\hat{\alpha}$ are related to each other by Onsager's relation: $\hat{\bar{\alpha}}=T\hat{\alpha}$. In the absence of charge current ($\mathbf{J}=0$), we have $\mathbf{E}=\hat{\sigma}^{-1}\hat{\alpha}\nabla T$. The Nernst coefficient $\vartheta$ can be derived to be
\begin{align}
\vartheta =   \frac{E_y}{(-dT/dx)} = \frac{\alpha_{xy}\sigma_{xx} - \alpha_{xx}\sigma_{xy}}{\sigma_{xx}^2 + \sigma_{xy}^2},
\label{nernst_eqn}
\end{align}
which is a function of thermoelectric tensor $\alpha_{ab}$ and charge conductivity tensor $\sigma_{ab}$. We will evaluate $\alpha_{ab}$ and $\sigma_{ab}$ using semi-classical Boltzmann treatment in the relaxation time approximation, accounting for an external magnetic field and a finite Berry curvature.

A non-zero Berry curvature in a Bloch Hamiltonian acts like a fictitious magnetic field in the momentum space \cite{Niu:2010}, which substantially modifies transport properties of the system, giving rise to anomalous behavior. Anomalous transport due to the Berry curvature has been crucial in understanding intrinsic Hall and Nernst conductivity in ferromagnetic materials\cite{Taguchi:2001, Fang:2003, Lee:2004, Niu:2006}. The Berry curvature for a Bloch Hamiltonian $H(\mathbf{k})$ is defined to be: $\Omega_{k_a}=\epsilon_{abc}\partial_{k_b}A_c(\mathbf{k})$, where $\mathbf{A}(\mathbf{k})$ is the Berry connection given by $\mathbf{A}(\mathbf{k})=\<u_{\mathbf{k}}|i\nabla_{\mathbf{k}}|u_{\mathbf{k}}\>$, for a Bloch eigenstate $|u_{\mathbf{k}}\>$. In the presence of Berry curvature $\mathbf{\Omega}_{\mathbf{k}}$, the semi-classical equation of motion for an electron takes the following form\cite{Niu:2006, Sundurum:1999}
\begin{eqnarray}
\mathbf{\dot{r}} = \frac{1}{\hbar} \frac{\partial\epsilon(\mathbf{k})}{\partial\mathbf{k}} + \frac{\mathbf{\dot{p}}}{\hbar}\times\mathbf{\Omega}_{\mathbf{k}},
\label{r_dot_eqn_1}
\end{eqnarray}
where $\mathbf{k}$ is the crystal momentum, $\epsilon(\mathbf{k})$ is the energy dispersion, and $\mathbf{p}=\hbar\mathbf{k}$. The first term in Eq.~\ref{r_dot_eqn_1} is the familiar relation between semi-classical velocity $\mathbf{\dot{r}}$ and the band energy dispersion $\epsilon(\mathbf{k})$. The second term is the anomalous transverse velocity term originating from $\mathbf{\Omega(\mathbf{k})}$. In the presence of electric and magnetic fields we have the standard relation: $\mathbf{\dot{p}} = e\mathbf{E} + e\mathbf{\dot{r}}\times\mathbf{B}$. These two coupled equations for $\mathbf{\dot{r}}$ and $\mathbf{\dot{p}}$ can be solved together to obtain~\cite{Duval:2006, Son:2012}
\begin{align}
&\mathbf{\dot{r}} = D(\mathbf{B},\Omega_{\mathbf{k}}) \[\mathbf{v}_{\mathbf{k}} + \frac{e}{\hbar} (\mathbf{E}\times\Omega_{\mathbf{k}}) + \frac{e}{\hbar}(\mathbf{v}_{\mathbf{k}}\cdot\Omega_{\mathbf{k}})\mathbf{B}\] \label{rdot_eqn}\\
&\mathbf{\dot{p}} = D(\mathbf{B},\Omega_{\mathbf{k}}) \[e\mathbf{E} + \frac{e}{\hbar} (\mathbf{v}_{\mathbf{k}}\times B) + \frac{e^2}{\hbar}(\mathbf{E}\cdot\mathbf{B})\mathbf{\Omega_{\mathbf{k}}}\], \label{pdot_eqn}
\end{align}
where $D(\mathbf{B},\Omega_{\mathbf{k}}) = (1+e (\mathbf{B}\cdot\Omega_{\mathbf{k}})/\hbar)^{-1}$. $D(\mathbf{B},\Omega_{\mathbf{k}})$ is also the prefactor which modifies the invariant phase space volume $d\mathbf{p}d\mathbf{x}\rightarrow D(\mathbf{B},\Omega_{\mathbf{k}})d\mathbf{p}d\mathbf{x}$, giving rise to a non-commutative mechanical model~\cite{Duval:2006}, because the Poisson brackets of coordinates is non-zero. For brevity of notation, we will sometimes omit showing the explicit dependence of $D(\mathbf{B},\Omega(\mathbf{k}))$ on $\mathbf{B}$ and $\Omega(\mathbf{k})$ and instead write just $D$. In Eq.~\ref{rdot_eqn} and Eq.~\ref{pdot_eqn}, we have also defined $\mathbf{v}_{\mathbf{k}}=\hbar^{-1}\partial\epsilon_{\mathbf{k}}/\partial\mathbf{k}$ to be the band-velocity. The second term in Eq.~\ref{rdot_eqn} gives rise to anomalous transport perpendicular to the applied electric field, while the third term gives rise to chiral magnetic effect. The third term in Eq.~\ref{pdot_eqn} (proportional to $\mathbf{E}\cdot\mathbf{B}$) is the source of chiral anomaly, triggering negative magnetoresistance. It has been shown recently that negative magnetoresistance can be derived using the semi-classical equations of motion employing Boltzmann transport\cite{Son:2013}. Other recent works have also developed a modified Boltzmann equation, taking into account Berry curvature and chiral anomaly effects\cite{Son:2012, Yamamato:2013, Chen:2013, Fiete:2014, Kim:2014}. In these works a linearized model of the WSM has been examined, i.e. a pair of  Dirac nodes topologically protected by chirality quantum numbers.

In this paper we solve the Boltzmann equation in the presence of the Berry curvature and chiral anomaly terms for a lattice model of a WSM. 
The steady state Boltzmann equation in the relaxation time approximation is given by
\begin{eqnarray}
(\mathbf{\dot{r}}\cdot \nabla_{\mathbf{r}} + \mathbf{\dot{k}}\cdot \nabla_{\mathbf{k}})f_{\mathbf{k}} =-\frac{f_{\mathbf{k}}-f_{eq}}{\tau},
\label{boltz_basic_eqn}
\end{eqnarray}
where $\tau$ is the scattering time, $f_{eq}$ is the equilibrium Fermi-Dirac distribution function, and $f_\mathbf{k}$ is the distribution function of the system in the presence of perturbations. The scattering time $\tau$ can in general be a function of the crystal momentum i.e. $\tau=\tau(\mathbf{k})$, but we shall treat it as independent of momentum for simplicity. 

We will first consider the case when $\mathbf{B}=0$ and derive the longitudinal and anomalous Hall conductivities. The linear response relations between the charge current and the applied fields dictate: 
\begin{align}
J_a = \sigma_{ab} E_b + \alpha_{ab} (-\nabla_b T)
\label{J_q_eqn}
\end{align}
The charge current in the presence of an electric field and a temperature gradient is given by\cite{Niu:2006}
\begin{align}
\mathbf{J} = &-e\int{[d\mathbf{k}] \left(\mathbf{v}_{\mathbf{k}} + \frac{e}{\hbar}\mathbf{E}\times\mathbf{\Omega_{\mathbf{k}}}\right) f_{\mathbf{k}}} \nonumber\\ &+ \frac{k_B e\mathbf{\nabla} T}{\hbar} \times \int{[d\mathbf{k}] \mathbf{\Omega_{\mathbf{k}}} s_{\mathbf{k}}}
\label{j1_eqn}
\end{align}
In the above expression, $[\mathbf{dk}]\equiv \frac{d^3k}{(2\pi)^3}$. The quantity $s_{\mathbf{k}} = -f_{eq}\log f_{eq} - ((1-f_{eq}) \log(1-f_{eq}))$ is entropy density for the electron gas. The first term in Eq.~\ref{j1_eqn} is the current in response to an applied electric field $\mathbf{E}$, also accounting for the transverse anomalous velocity acquired by an electron wave-packet due to $\mathbf{\Omega(\mathbf{k})}$. The second term is the anomalous response to the temperature gradient $\nabla T$, which can be obtained using the semiclassical wavepacket methods taking into account the orbital magnetization of the carriers arising from the finite spread of the wavefunction~\cite{Niu:2006}. It can also be derived by first calculating the transverse heat current in response to an electric field and then using Onsager's relation~\cite{Zhang:2008}.
The heat current $\mathbf{Q}$ takes the following form after accounting for both normal and anomalous contributions\cite{Niu:2006, Qin:2011, Yokoyama:2011, Bergman:2010}
\begin{align}
&\mathbf{Q} = \int{[d\mathbf{k}] (\epsilon_{\mathbf{k}}-\mu)\mathbf{v}_{\mathbf{k}}f_{\mathbf{k}}} + \frac{e}{\beta\hbar}\int{[d\mathbf{k}]\(\mathbf{E}\times\mathbf{\Omega_{\mathbf{k}}}\)s_{\mathbf{k}}}\nonumber \\ &  + \frac{k_B\nabla T}{\beta\hbar}\times\int{[d\mathbf{k}] \mathbf{\Omega}_{\mathbf{k}}\(\frac{\pi^2}{3}f_{eq}+\beta^2(\epsilon-\mu)^2f_{eq}\)}\nonumber \\
& - \frac{k_B\nabla T}{\beta\hbar}\times\int{[d\mathbf{k}] \mathbf{\Omega}_{\mathbf{k}}\(\mbox{ln}(1+e^{-\beta(\epsilon_{\mathbf{k}}-\mu)^2})+2\mbox{Li}_2(1-f_{eq})\)},
\label{Q1_eqn}
\end{align}
where $\mbox{Li}_2(z)$ is the polylogarithmic function of order 2, which is generally defined as $\mbox{Li}_s(z)=\sum\limits_{k=1}^{\infty}{\frac{z^k}{k^s}}$ for an arbitrary complex order $s$, for a complex argument $|z|<1$. The first term in Eq.~\ref{Q1_eqn} is the standard contribution to the heat current in the absence of Berry curvature. The second term is the Berry curvature mediated transverse response to electric field $\mathbf{E}$ which can be understood by the following simple argument: in the presence of the Berry curvature and the electric field, the electron velocity acquires the additional anomalous term $e\mathbf{E}\times\mathbf{\Omega_{\mathbf{k}}}$. Multiplying this velocity by the entropy density of the electron gas, we obtain this contribution to the transverse heat current~\cite{Zhang:2008}. From Eq.~\ref{thermal} we can  write the transverse response of $\mathbf{J}$ and $\mathbf{Q}$ on the applied temperature gradient and electric field respectively as: $J_x=\alpha_{xy}\nabla_y T$, and $Q_x=\bar{\alpha}_{xy} E_y$. Comparing the coefficients $\alpha_{xy}$ and $\bar{\alpha}_{xy}$ from Eq.~\ref{j1_eqn} and Eq.~\ref{Q1_eqn}, it is easy to note that they obey Onsager's relation: $\bar{\alpha}_{xy}=T\alpha_{xy}$, as expected. From Eq.~\ref{thermal}, the anomalous response $\mathbf{Q}$ on an applied temperature gradient can be written as: $Q_x=l_{xy}\nabla_y T$. The quantity $l_{xy}$ in Eq.~\ref{thermal} can be calculated as: $l_{xy}=-k_B^2T c_2/\hbar$, where~\cite{Bergman:2010}
\begin{align}
c_n=\int{[d\mathbf{k}]\Omega_z\int\limits_{\epsilon-\mu}^{\infty} d\epsilon(\beta\epsilon)^n \frac{\partial f_{eq}}{\partial\epsilon}}
\label{c_n_eqn}
\end{align} 
The energy integral in Eq.~\ref{c_n_eqn} reduces to the following for $n=2$ \cite{Bergman:2010, Yokoyama:2011}.
\begin{align}
&\int\limits_{\epsilon-\mu}^{\infty} d\epsilon(\beta\epsilon)^2 \frac{\partial f_{eq}}{\partial\epsilon}=\frac{\pi^2}{3}f_{eq}+\beta^2(\epsilon-\mu)^2f_{eq}\nonumber\\
&-\(\mbox{ln}(1+e^{-\beta(\epsilon_{\mathbf{k}}-\mu)^2})+2\mbox{Li}_2(1-f_{eq})\)
\label{energy_intg_eqn}
\end{align}
Eq.~\ref{energy_intg_eqn} and Eq.~\ref{c_n_eqn} combined with $l_{xy}=-k_B^2T c_2/\hbar$ give the last two $\nabla T$ dependent terms in Eq.~\ref{Q1_eqn}.

Keeping only linear order dependence on the applied field $\mathbf{E}$ and $\nabla T$, the following Ansatz is assumed for the distribution function, $f_{\mathbf{k}}$~\cite{Ashcroft} which is a solution to the steady state Boltzmann equation (Eq~\ref{boltz_basic_eqn}) :
\begin{eqnarray}
f_{\mathbf{k}} = f_{eq} + \tau \left(-\frac{\partial f_{eq}}{\partial \epsilon}\right) \mathbf{v}_{\mathbf{k}}\cdot \left( -e\mathbf{E} + \frac{\epsilon-\mu}{T} (-\nabla T)\right),
\label{f_ansatz1}
\end{eqnarray}
We will further assume that the electric field and temperature gradient have non-zero components only along the $x$ direction. Substituting for $f_{\mathbf{k}}$ from Eq.~\ref{f_ansatz1} in Eq.~\ref{j1_eqn} for the current $\mathbf{J}$, and comparing the resulting expression with Eq.~\ref{J_q_eqn}, the longitudinal components of the conductivity tensors $\sigma_{ab}$ and $\alpha_{ab}$ can be easily read to be
\begin{align}
&\sigma_{xx} = e^2\int{[d\mathbf{k}] v_x^2 \tau \left(-\frac{\partial f_{eq}}{\partial\epsilon}\right)},\label{sxx_eq_1}\\
&\alpha_{xx} = -\frac{e}{T}\int{[d\mathbf{k}] v_x^2 \tau (\epsilon-\mu) \left(-\frac{\partial f_{eq}}{\partial\epsilon}\right)}, \label{axx_eq_1}
\end{align}
where $v_x\equiv\hbar^{-1}\partial \epsilon_{\mathbf{k}}/\partial k_x$. The transverse components are:
\begin{align}
&\sigma_{yx} = \frac{e^2}{\hbar} \int{[d\mathbf{k}] \mathbf{\Omega} f_{eq}} ,\label{syx1}\\
&\alpha_{yx} = {\frac{k_Be}{\hbar} \int{[d\mathbf{k}] \mathbf{\Omega_{\mathbf{k}}} s_{\mathbf{k}}}},
\label{ayx1}
\end{align}
which are purely anomalous because we have assumed $\mathbf{E}$ and $\nabla T$ are applied along the $x$ direction, and there is no magnetic field. 

We now discuss the case of a finite magnetic field. We consider a particular configuration relevant for the experiments measuring Nernst coefficient i.e. $\nabla T = \nabla_x T\hat{x}$, $\mathbf{B} = B \hat{z}$, and $\mathbf{E}=0$, although the approach will work for other configurations also, like the parallel setup discussed in Section V. The Boltzmann equation (Eq.~\ref{boltz_basic_eqn} takes the following form after making substitutions for $\dot{\mathbf{r}}$ and $\dot{\mathbf{p}}$ from Eq.~\ref{rdot_eqn} and~\ref{pdot_eqn}).
\begin{align}
&v_x\tau\nabla_x T\frac{\epsilon-\mu}{T}\left(-\frac{\partial f_{eq}}{\partial\epsilon}\right) + \frac{eB}{\hbar}\left(-v_x\frac{\partial}{\partial k_y} + v_y\frac{\partial}{\partial k_x}\right)f_{\mathbf{k}}\nonumber \\
&= -\frac{f_{\mathbf{k}}-f_{eq}}{D(B,\mathbf{\Omega}_{\mathbf{k}})\tau}
\label{boltz_eqn_5}
\end{align}
The following Ansatz is chosen for the distribution $f_{\mathbf{k}}$ which also accounts for correction factor ($\mathbf{\Lambda}$) due to a finite magnetic field.
\begin{align}
f_{\mathbf{k}}= f_{eq} - \(D\tau v_x \nabla_x T \frac{\epsilon-\mu}{T}-\mathbf{v}\cdot\mathbf{\Lambda}\) \left(-\frac{\partial f_{eq}}{\partial\epsilon}\right)
\label{f_ansatz2}
\end{align}
The Boltzmann equation (Eq.~\ref{boltz_eqn_5}) thus becomes
\begin{align}
&\frac{eB}{\hbar}\(v_y\frac{\partial}{\partial{k_x}} - v_x\frac{\partial}{\partial k_y} \) (-D\nabla_x T \frac{\epsilon-\mu}{T}v_x\tau + \mathbf{v}\cdot\mathbf{\Lambda}) \nonumber \\
&= -\frac{\mathbf{v}\cdot\mathbf{\Lambda}}{\tau}
\end{align}
Imposing the condition that this equation must be valid for all values of $\mathbf{v}$, we find that $\Lambda_z=0$, and the equation can be simplified to:
\begin{align}
&eB\nabla_x T\frac{\epsilon-\mu}{T} D\tau \(\frac{v_x}{m_{xy}} - \frac{v_y}{m_{xx}}\) + eB\( \frac{v_y\Lambda_x}{m_{xx}} - \frac{v_x\Lambda_y}{m_{yy}}\) \nonumber \\
&=-v_x\Lambda_x\(-\frac{eB}{m_{xy}} + \frac{1}{D\tau}\) -v_y\Lambda_y\(\frac{eB}{m_{xy}} + \frac{1}{D\tau}\)
\end{align}
In order to solve the above equation, we introduce complex variables $V=v_x+iv_y$, and $\Lambda=\Lambda_x - i\Lambda_y$, and rewrite the equation in the following manner
\begin{align}
\mbox{Re}\[eB\tau D \nabla_x T\frac{\epsilon-\mu}{T} V\(\frac{1}{m_{xy}}+\frac{i}{m_{xx}}\)\]\nonumber \\=\mbox{Re}\[V\Lambda\(\frac{ieB}{m_{xx}}-\frac{1}{D\tau}\)+\frac{eBV\Lambda^*}{m_{xy}}\],
\label{boltz_eqn_4}
\end{align}
where $m_{ij}^{-1} = \hbar^{-2}\partial^2E(\mathbf{k})/\partial k_i\partial k_j$ is the inverse band-mass tensor and Re($z$) stands for the real part of $z$. Eq.~\ref{boltz_eqn_4} can be solved for $\Lambda$:
\begin{widetext}
\begin{align}
\Lambda_x = eB\tau D(B,\mathbf{\Omega}_{\mathbf{k}})\nabla_x T\frac{\epsilon-\mu}{T}\frac{\[\frac{v_x}{m_{xy}}-\frac{v_y}{m_{xx}}\]\[-\frac{eBv_y}{m_{xx}}+\frac{eBv_x}{m_{xy}}-\frac{v_x}{D\tau}\] + \[\frac{v_x}{m_{xx}}+\frac{v_y}{m_{xy}}\]\[\frac{eBv_x}{m_{xx}}-\frac{eBv_y}{m_{xy}}-\frac{v_y}{D\tau}\]}{\[-\frac{eBv_y}{m_{xx}}+\frac{eBv_x}{m_{xy}}-\frac{v_x}{D\tau}\]^2 + \[\frac{eBv_x}{m_{xx}}-\frac{eBv_y}{m_{xy}}-\frac{v_y}{D\tau}\]^2}
\label{lx1_eqn}
\end{align}
\begin{align}
\Lambda_y = eB\tau D(B,\mathbf{\Omega}_{\mathbf{k}})\nabla_x T\frac{\epsilon-\mu}{T}\frac{\[\frac{v_x}{m_{xy}}-\frac{v_y}{m_{xx}}\]\[-\frac{eBv_y}{m_{xy}}+\frac{eBv_x}{m_{xx}}-\frac{v_y}{D\tau}\] - \[\frac{v_x}{m_{xx}}+\frac{v_y}{m_{xy}}\]\[-\frac{eBv_y}{m_{xx}}+\frac{eBv_x}{m_{xy}}-\frac{v_x}{D\tau}\]}{\[-\frac{eBv_y}{m_{xx}}+\frac{eBv_x}{m_{xy}}-\frac{v_x}{D\tau}\]^2 + \[\frac{eBv_x}{m_{xx}}-\frac{eBv_y}{m_{xy}}-\frac{v_y}{D\tau}\]^2}
\label{ly1_eqn}
\end{align}
\end{widetext}
For convenience of notation, we rewrite $\Lambda_x$ and $\Lambda_y$ as: $\Lambda_i=\tau \nabla_x T\frac{\epsilon-\mu}{T} c_i$, incorporating into $c_i$ the remaining factors apart from $\tau \nabla_x T\frac{\epsilon-\mu}{T}$ of Eq.~\ref{lx1_eqn} and Eq.~\ref{ly1_eqn}. Using Eq.~\ref{f_ansatz2} and the results for $\Lambda_x$ and $\Lambda_y$, we can now explicitly write the distribution function $f_{\mathbf{k}}$ as:
\begin{align}
f_{\mathbf{k}} = f_{eq} - \left(\tau\nabla_x T \frac{\epsilon-\mu}{T} \left(\frac{\partial f_{eq}}{\partial\epsilon}\right)\right) \left((c_x-D) v_x + c_y v_y \right)
\label{f_2}
\end{align}
The expression for the charge current $\mathbf{J}$, in the presence of $\mathbf{B}$ and $\Omega_{\mathbf{k}}$, is also modified by the factor $D(\mathbf{B},\Omega_{\mathbf{k}})$~\cite{Fiete:2014, Niu:2006}, as we pointed out earlier that $D(\mathbf{B},\Omega_{\mathbf{k}})=(1+e (\mathbf{B}\cdot\Omega_{\mathbf{k}})/\hbar)^{-1}$ is the multiplicative factor which alters the phase space volume locally. 
\begin{align}
\mathbf{J} = -e\int{[d\mathbf{k}]D^{-1}\mathbf{\dot{r}} f} + \frac{k_B e\mathbf{\nabla} T}{\hbar} \times \int{[d\mathbf{k}] \mathbf{\Omega_{\mathbf{k}}} s_{\mathbf{k}}}
\label{J2}
\end{align}
Substituting Eq.~\ref{f_2} in Eq.~\ref{J2} and again comparing with the linear response relations in Eq.~\ref{thermal}, the thermoelectric tensor $\alpha_{ij}$ can be solved to: 
\begin{align}
\alpha_{xx} &= e\int{[d\mathbf{k}] v_x^2\left(\tau\frac{\epsilon-\mu}{T} \left(-\frac{\partial f_{eq}}{\partial\epsilon}\right) \left(c_x-D \right) \right)}
\label{axx2}\\
\alpha_{yx} &= e\int{[d\mathbf{k}] (v_y^2 c_y + (c_x-D)v_xv_y) \left(\tau\frac{\epsilon-\mu}{T} \left(-\frac{\partial f_{eq}}{\partial\epsilon}\right) \right)}\nonumber \\
&+ \frac{k_B e}{\hbar} \int{[d\mathbf{k}] {\Omega_z} s_{\mathbf{k}}}
\label{ayx2}
\end{align}
The temperature dependence of the zero-field anomalous contribution in Eq.~\ref{ayx2} is hidden in the entropy density $s_{\mathbf{k}}$ of the electron gas. Similarly, the electrical conductivity components (transverse and longitudinal) are obtained to be:
\begin{align}
\sigma_{xx} &= e^2\int{[d\mathbf{k}] v_x^2\tau \left(-\frac{\partial f_{eq}}{\partial\epsilon}\right) \left(c_x-D \right)}
\label{sxx2}\\
\sigma_{yx} &= {e^2}\int{[d\mathbf{k}] (v_y^2 c_y\ + v_xv_y(c_x-D)) \tau\left(-\frac{\partial f_{eq}}{\partial\epsilon}\right)}\nonumber \\
&+ \frac{e^2}{\hbar} \int{[d\mathbf{k}] {\Omega_z}f_{eq}}
\label{syx2}
\end{align}
As a good check of our calculation we also recover the results for $\sigma_{ab}$ and $\alpha_{ab}$ found in Ref.~\onlinecite{Fiete:2014} where the tensorial nature of $m_{ij}$ is ignored and $m=\hbar\mu/v_F^2$. The transverse components, i.e. Eq.~\ref{ayx2} and ~\ref{syx2}, are a sum of two terms: the first term captures the effect of a finite $\mathbf{B}$ which is further modified by the Berry curvature $\Omega_{\mathbf{k}}$, due to the factors $c_y$, $c_x$ and $D$ which are non-trivial functions of the Berry curvature. The first terms in Eq.~\ref{ayx2} and Eq.~\ref{syx2} depend on the scattering time $\tau$, and we call these as `modified' $B$-dependent Hall conductivities (because they are modified due to the Berry curvature). The Berry curvature also alters the expressions for longitudinal conductivities given in Eq.~\ref{axx2} and Eq.~\ref{sxx2} because of the factor $c_x-D$. In the limits when $\Omega_{\mathbf{k}}\rightarrow 0$, the factor $c_y\rightarrow \omega\tau$ for a quadratic band dispersion, upto linear order in $\mathbf{B}$,  (where $\omega=eB/m$ is the cyclotron frequency). In the same limit, the factor $c_x-D\rightarrow -1$ upto zeroth order in $\mathbf{B}$, thus yielding the standard expression for $\sigma_{xx}$ and $\alpha_{xx}$ given in Eq.~\ref{sxx_eq_1} and ~\ref{axx_eq_1}. In contrast, the second term in Eq.~\ref{ayx2} and ~\ref{syx2}, which is Berry curvature dependent persists in the absence of a magnetic field, and is a purely anomalous contribution. We shall roughly examine the limit in which the factor $D(\mathbf{B},\Omega_{\mathbf{k}}) = (1+e (\mathbf{B}\cdot\Omega_{\mathbf{k}})/\hbar)^{-1}$ significantly deviates from 1. Defining $\mathbf{k}=2\pi\mathbf{K}/a$, where $a$ is the lattice constant, and $\mathbf{K}$ is dimensionless, $e (\mathbf{B}\cdot\Omega_{\mathbf{k}})/\hbar=a^2\Omega_{\mathbf{K}}/l_B^2$, where $l_B$ is the magnetic length and $\Omega_{\mathbf{K}}$ is dimensionless. For a magnetic field of 1T, $a\sim 2\AA$, $e(\mathbf{B}\cdot\Omega_{\mathbf{k}})/\hbar \sim 10^{-4}\Omega_{K}$. The Berry curvature for a single linearly dispersing Weyl node centered at the origin is ${\Omega}_{\mathbf{K}}=\mathbf{K}/4|\mathbf{K}|^3$, thus $D(B,{\Omega}_{\mathbf{k}})\approx 1$ for low magnetic fields and away from band touching point ($\mathbf{K}_0=0$) in the momentum space (i.e. approximately $|\mathbf{K}|\gg 0.01$ in this case). This is expected, as qualitatively one understands that the effect of Berry curvature peaks when the energy band gap $E_g\rightarrow 0$. When the effects of Berry curvature can be neglected, the following standard expressions are derived from Eq.~\ref{ayx2} and~\ref{syx2} for Hall conductivities, keeping terms only upto linear order in $\mathbf{B}$:
\begin{align}
&\sigma_{xy} = \frac{-e^3\tau^2B}{\hbar} \int{[d\mathbf{k}] \(\frac{\partial f_0}{\partial\epsilon}\)\(\frac{v_x^2\partial^2\epsilon}{\partial k_y^2}-\frac{v_xv_y\partial^2\epsilon}{\partial k_x\partial k_y}\)}
\label{Hall_eqn_1}
\end{align}
\begin{align}
&\alpha_{xy} = \frac{-e^3\tau^2B}{T\hbar} \int{[d\mathbf{k}] (\epsilon-\mu)\(\frac{\partial f_0}{\partial\epsilon}\)\(\frac{v_x^2\partial^2\epsilon}{\partial k_y^2}-\frac{v_xv_y\partial^2\epsilon}{\partial k_x\partial k_y}\)}
\label{Hall_eqn_2}
\end{align}
In Section III, we will use the formula obtained for $\alpha_{ab}$ and $\sigma_{ab}$ to calculate the Nernst coefficient in Eq.~\ref{nernst_eqn}, first analytically for a simple Dirac and Weyl linearized Hamiltonian, and then numerically in Section IV, for a lattice model of Weyl fermions. 

\section{Nernst response in linearized model Dirac and Weyl systems}
In this section, we will concern ourselves with the Nernst response of a linearized spectrum of Dirac and Weyl systems. We examine the magnetic field dependent transverse conductivities ($\alpha_{xy}$ and $\sigma_{xy}$) for a linearly dispersing Dirac node, which are analytically tractable using the Boltzmann approach. We repeat the procedure for a pair of Weyl nodes taking into consideration the Berry flux modification of the normal $B$-dependent conductivities.
\subsection{Nernst effect in a linearized Dirac Hamiltonian}
As a warm up, we discuss the Nernst response of a single Dirac cone, with linear dispersion $\epsilon_{\mathbf{k}}=\pm\hbar v_F \mathbf{k}$, where $\mathbf{k}=\sqrt{k_x^2 + k_y^2 + k_z^2}$. The density of states for a single Dirac/Weyl node with unbounded linear dispersion (taking into account spin degeneracy) is given by
\begin{eqnarray}
\rho(E) = \frac{1}{\pi^2} \frac{E^2}{(\hbar v_F)^3}
\end{eqnarray}
The density of states vanishes at the Dirac node, which gives rise to many unusual properties. Eq.~\ref{sxx2} for the longitudinal conductivity, without the Berry curvature term, reduces to Eq.~\ref{sxx_eq_1}, which can be employed to analytically deduce the zero temperature conductivity for a Dirac Hamiltonian to be
\begin{align}
\sigma_{xx} = \frac{e^2}{6\pi^2} \frac{\tau\mu^2}{\hbar^3 v_F},
\label{sxx_dirac}
\end{align}
where we have assumed the scattering time $\tau$ to be a phenomenological parameter independent of energy or momentum. For transverse magneto-conductivity, for a weak magnetic field $\mathbf{B}=B\hat{z}$, we derive the Hall conductivity $\sigma_{xy}$ for a linearized Dirac Hamiltonian using Eq.~\ref{Hall_eqn_1} to be
\begin{eqnarray}
\sigma_{xy} = \frac{e^3B\tau^2}{6\pi^2}\frac{v_F\mu}{\hbar^4}
\label{sxy_dirac}
\end{eqnarray}
We note that $(\sigma_{xy}/\sigma_{xx}) = \omega\tau = eB\tau/m$, where $m$ is the band mass near the Fermi surface given by $m=\hbar\mu/v_F^2$. At low temperatures, the thermoelectric tensor $\alpha_{ab}$ is related to the derivative of $\sigma_{ab}$ via the Mott relation~\cite{Ziman}
\begin{align}
\alpha_{ab} = -\frac{\pi^2}{3}\frac{k_B^2 T}{e}\frac{\partial\sigma_{ab}}{\partial\mu}
\label{Mott_eqn}
\end{align}
Combining the Mott relation with Eq.~\ref{nernst_eqn}, the Nenrst coefficient becomes (when $\sigma_{xx}\gg\sigma_{xy}$)
\begin{eqnarray}
\vartheta = -\frac{\pi^2}{3}\frac{k_B^2 T}{e} \frac{\partial \Theta_H}{\partial \mu} = -{\pi^2k_B^2 T}\frac{B\tau v_F^2}{3\hbar\mu^2},
\label{nernst_mott_eqn}
\end{eqnarray}
where $\Theta_H=\sigma_{xy}/\sigma_{xx}$ is the Hall angle. However Eq.~\ref{nernst_mott_eqn} is valid only when $\mu\neq 0$, and $\vartheta$ does not diverge for $\mu=0$ as we shall see shortly. 

The scattering time in Boltzmann conductivity is sensitive to the type of impurities in the system. For neutral short range or point-like impurities, the scattering time $\tau_s$ is given by:\cite{Sarma:2015}
\begin{align}
\frac{1}{\tau_s} = \frac{n_s V_0^2k_F^2}{3\pi\hbar^2 v_F},
\label{tau_s_eqn}
\end{align}
where $n_s$ is the density of the impurities, $V_0$ is the strength of the impurity potential, and $k_F$ is the Fermi wave-vector. Considering Thomas-Fermi (TF) screening, the scattering time for long-range ionic impurities at zero temperature is given by\cite{Sarma:2015}
\begin{equation}
\frac{1}{\tau_c} = 4\pi\alpha^2 n_c\frac{v_F}{k_F^2} I_t(q_0)
\label{tau_c_eqn}
\end{equation}
In the above expression, $n_c$ is the density of charged impurities, $\alpha=e^2/\kappa\hbar v_F$ is the fine-structure constant, $q_0=q_{TF}/2k_F$, where $q_{TF}$ is the Thomas-Fermi wave-vector, and $I_t(x)=(x^2 + 1/2)\log(1+1/x^2)-1$. The total scattering time is given by Matthiesen's rule
\begin{align}
\frac{1}{\tau} = \frac{1}{\tau_s} + \frac{1}{\tau_c}
\label{matt_rule_eqn}
\end{align}
For a linear Dirac Hamiltonian, $k_F=\mu/\hbar v_F$, therefore the scattering time expression take the following form:
\begin{align}
\frac{1}{\tau_c} &= \frac{n_c4\pi\alpha^2 v_F^3\hbar I_t(q_0)}{\mu^2} \label{tau_c_1_eqn}\\
\frac{1}{\tau_s} &= \frac{n_sV_0^2\mu^2}{3\pi\hbar^3v_F} \label{tau_s_1_eqn}
\end{align}
From Matthiesen's rule, it is evident that the shorter time scattering process ($\tau_s$ or $\tau_c$) will dominate the carrier transport, therefore near $\mu=0$, the scattering from ionic impurities will primarily determine the conductivity, and for $\mu\gg0$, it is scattering from the neutral point-like impurities that govern charge transport. For an arbitrary value of $\mu$, the following expression for scattering time $\tau$ can be written, using expressions in Eq.~\ref{matt_rule_eqn},~\ref{tau_c_1_eqn}, and~\ref{tau_s_1_eqn}
\begin{eqnarray}
\frac{1}{\tau} = \frac{1}{\tau_0} \left(\frac{1}{\mu^2}(1+x\mu^4) \right),
\label{total_tau_eqn_1}
\end{eqnarray}
where $\tau_0$ and $x$ are constants depending on the coefficients of $\mu^2$ and $1/\mu^2$ in Eq.~\ref{tau_c_1_eqn} and~\ref{tau_s_1_eqn}, whose exact form is lengthy and not illuminating for our discussion. Using this expression for the total scattering time $\tau$ in Eq.~\ref{sxx_dirac}, we obtain:
\begin{align}
\sigma_{xx} = \frac{e}{6\pi^2}\frac{\tau_0\mu^4}{v_F\hbar^3(1+x\mu^4)}
\label{sxx_dirac_2}
\end{align}
Similarly, from Eq.~\ref{sxy_dirac}, we have
\begin{align}
\sigma_{xy}=\frac{e^2}{6\pi^2}\frac{eBv_F\tau_0^2\mu^5}{\hbar^4(1+x\mu^4)^2}
\label{sxy_dirac_2}
\end{align}
The Hall angle near $\mu=0$ reduces to:
\begin{align}
\Theta_H=eB\mu v_F^2\tau_0/\hbar,
\label{Hall_at_mu_0}
\end{align}
and thus from Eq.~\ref{nernst_mott_eqn}, the Nernst coefficient $\vartheta_0$ at $\mu=0$ is given by:
\begin{equation}
\vartheta_0 = -\frac{\pi^2}{3}\frac{k_B^2 T}{e}\frac{eBv_F^2\tau_0}{\hbar}
\end{equation}
Alternatively, the same conclusion can be reached by using the equation for the Nernst coefficient i.e. Eq.~\ref{nernst_eqn}, and Eq.~\ref{Mott_eqn},~\ref{sxx_dirac_2}, and~\ref{sxy_dirac_2}, by substituting the exact expressions instead of using the Mott relation. Far away from the Dirac point at $\mu=0$, the charge conductivities are
\begin{align}
\sigma_{xx} &\sim \frac{e^2}{6\pi^2} \frac{\tau_0}{v_F\hbar^3}x\(1-\frac{1}{x\mu^4}\) \label{sigmaxx_dirac_3}\\
\sigma_{xy} &\sim \frac{e^2}{6\pi^2}\frac{eBv_F\tau_0^2}{\hbar^4x^2\mu^3}
\end{align}
The Hall angle in this case no longer varies linearly with the Fermi energy (as in the case near $\mu=0$), but is instead given by $\Theta_H=eBv_F^2\tau_0/\hbar x^2\mu^3$. The Nernst coefficient for $\mu\gg 0$ becomes
\begin{align}
\vartheta \sim \frac{\pi^2k_B^2 T}{e}\frac{eBv_F^2\tau_0}{\hbar x^2\mu^4},
\end{align}
which approaches zero as $\mu$ is increased asymptotically, as expected from Fermi liquid theory, where the Nernst coefficient vanishes because of Sondheimers cancellation.\cite{Ong:2001}
\begin{figure}[h]
\includegraphics[scale=.2]{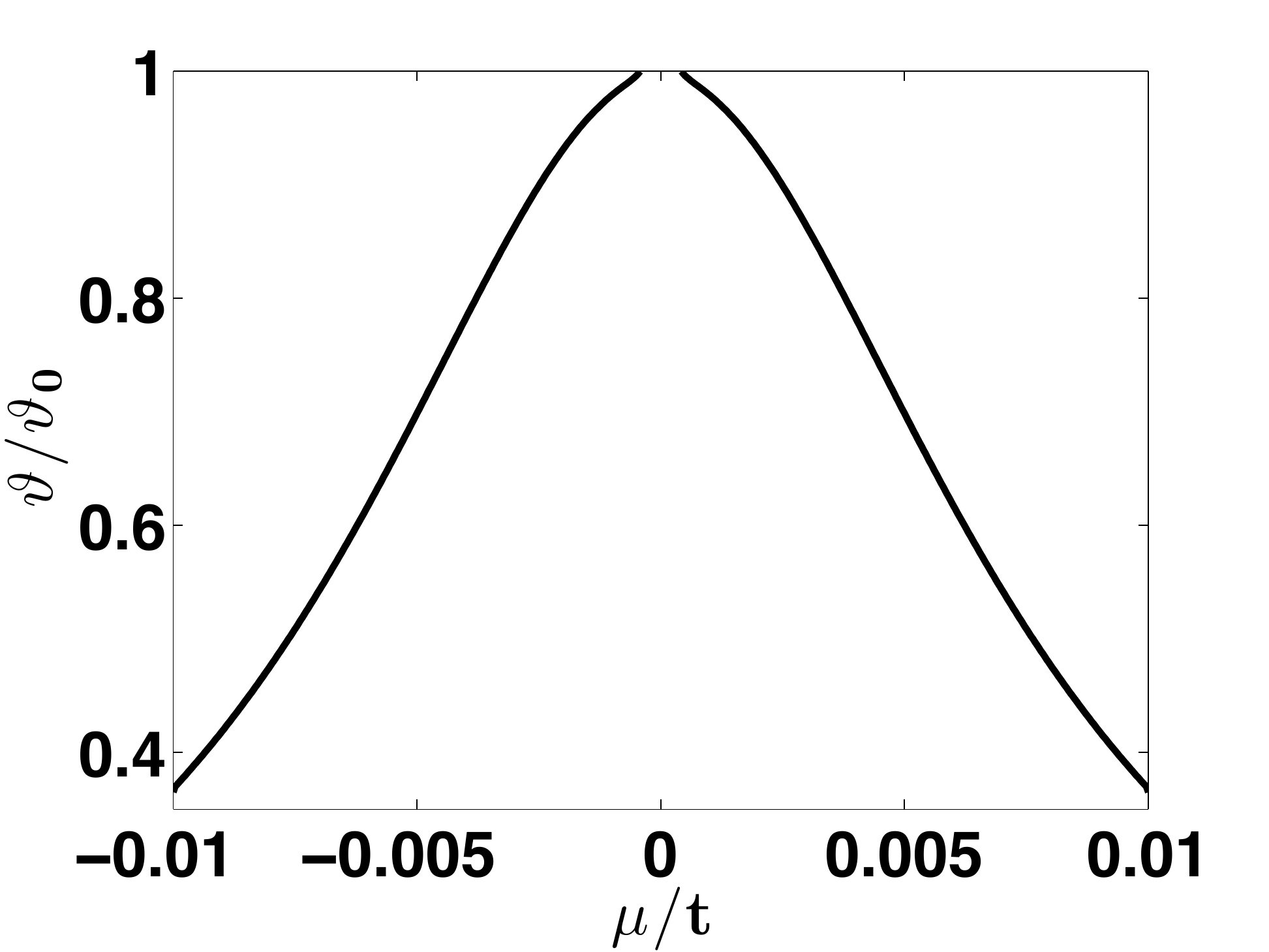}
\includegraphics[scale=.22]{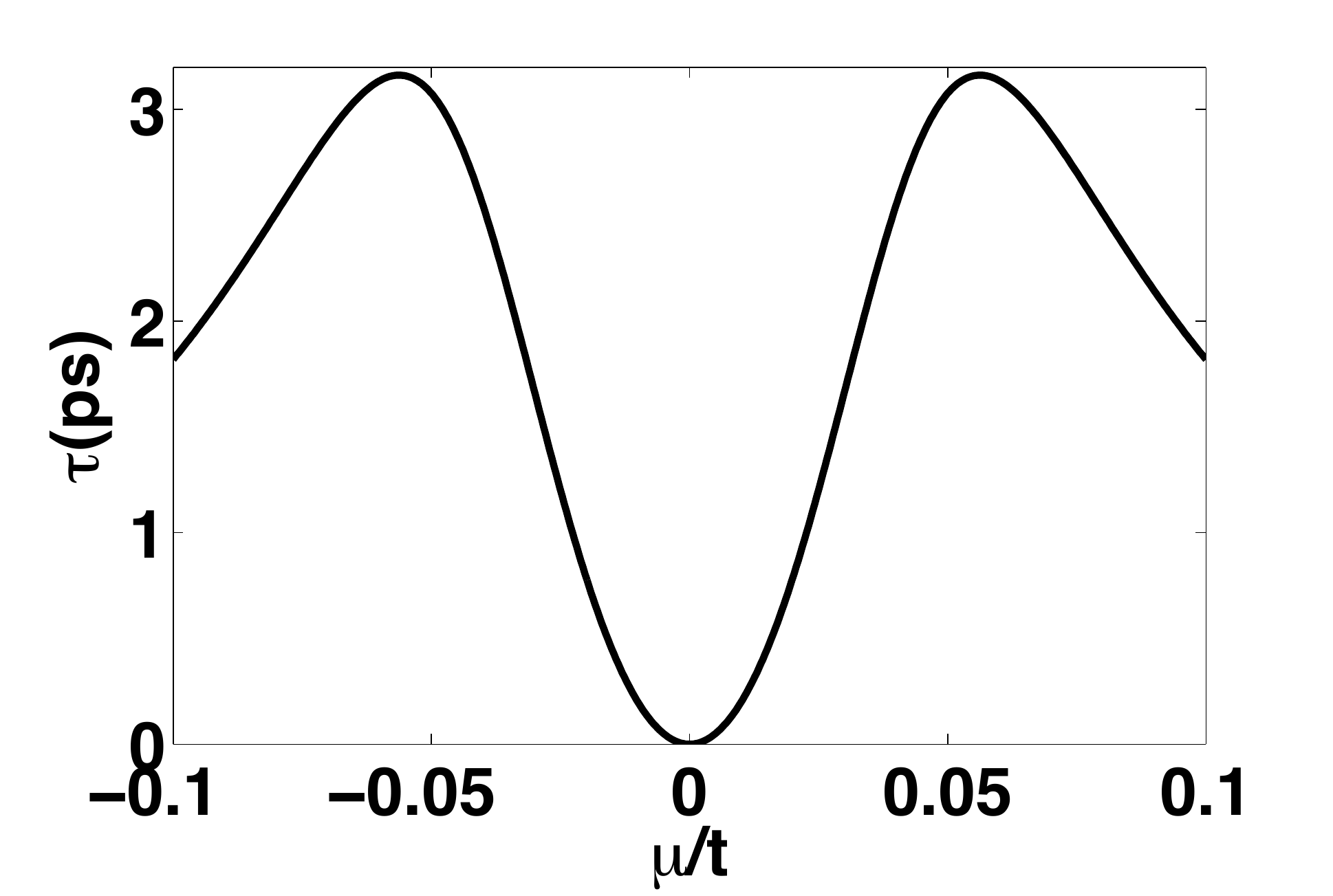}
\caption{\textit{Left:} Plot of the Nernst coefficient $\vartheta$ as a function of $\mu/t$ for a single linearized Dirac node whose spectrum is bounded at $\pm$ $3.5t$. The energy parameter $t$ is chosen to be $t=0.1$ $eV$. Here $\vartheta_0$ indicates the value of the Nernst coefficient at $\mu=0$. \textit{Right:} Chosen scattering time for the calculation in $ps$ as given by Eq.~\ref{total_tau_eqn_1}.}
\label{nu_vs_mu_dirac}
\end{figure}
The sign of the Nernst coefficient $\vartheta$ does not change with the sign of $\mu$ and is thus an even function of $\mu$. This is because both $\sigma_{xy}\alpha_{xx}$ and $\sigma_{xx}\alpha_{xy}$, which appear in the numerator of the expression for $\vartheta$ in Eq.~\ref{nernst_eqn} do not depend on sgn($\mu$). The plot in Figure~\ref{nu_vs_mu_dirac} displays Nernst coefficient for a linear Dirac node obtained using Eq.~\ref{nernst_eqn},~\ref{sxx_eq_1},~\ref{Hall_eqn_1}, and~\ref{Mott_eqn}. We have provided a physical ultra-violet cut-off to the low energy spectrum at $\epsilon=\pm3.5t$ with $t=0.1$ $eV$, $v_F\sim 10^5$ $m/s$ $T=40K$, and $B=1T$. The chosen scattering time $\tau$ used for this calculation, and its scaling with $\mu$ is also shown in Figure~\ref{nu_vs_mu_dirac}. A regularized lattice model will however smoothly bound the dispersion in the Brillouin zone. Section IV will be devoted to evaluating the Nernst response of a WSM Hamiltonian defined on a lattice.
\subsection{Nernst effect for a pair of linearized Weyl fermions}
\begin{figure}[h]
\includegraphics[scale=.35]{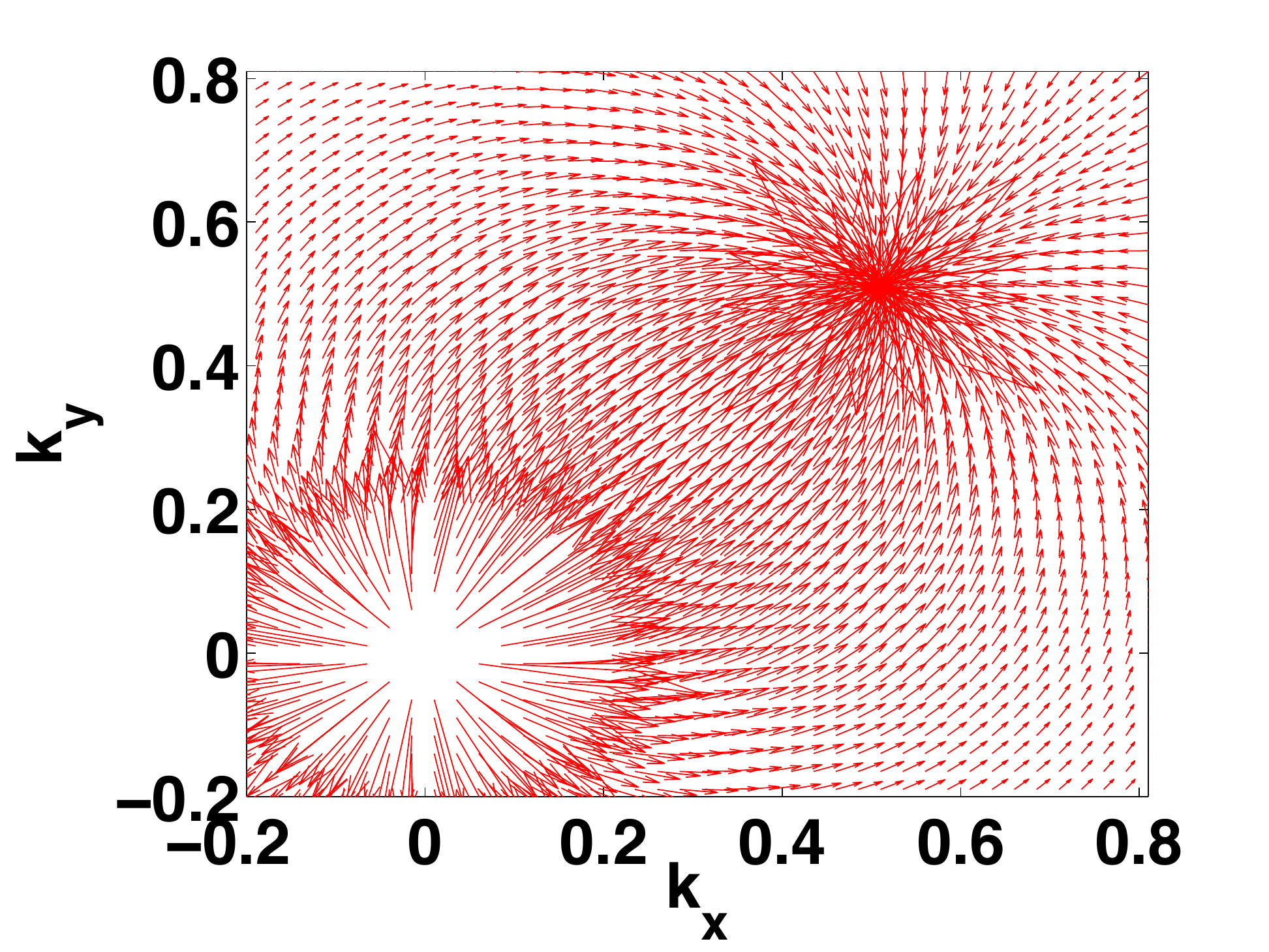}
\caption{(color online) Plot of Berry curvature of a Weyl semimetal in the $k_z=0$ plane, where the node separation is $\mathbf{k}_0=(0.5,0.5,0)$. At the origin is a Weyl node of positive chirality, which acts as a source of Berry flux indicated by outgoing arrows, which flow towards $\mathbf{k}_0$, which acts as a sink of Berry flux indicated by incoming arrows.}
\label{weyl_2_nodes_berry}
\end{figure}

A single Dirac node can be visualized as two Weyl nodes, which are topologically protected by chirality quantum number of opposite sign, coinciding with each other in energy-momentum space. Therefore the net flux of the Berry curvature, and henceforth the net chirality vanishes for a single Dirac node, resulting in a zero anomalous response for both the charge and thermoelectric conductivity. As a result, no anomalous Nernst response is also expected in a linear Dirac Hamiltonian. Using an external perturbation, a single band-touching point in a Dirac cone can be shifted into a pair of isolated Weyl points possessing opposite chirality quantum numbers. The external perturbation must break either time-reversal symmetry or inversion symmetry, and also lifts-up the degeneracy of the Dirac spinor. Our discussion will be centered upon the assumption that time-reversal symmetry is violated, which can be achieved using magnetic field as a perturbation. Also the Weyl points are assumed to occur at the same energy, thus there is no chiral chemical potential. One can also construct inversion asymmetric and TR invariant models of a WSM, but they are not of interest to us here because the anomalous Hall and Nernst response vanishes as the vector sum $\mathbf{k}_0$ of the node separation becomes zero. Figure~\ref{weyl_2_nodes_berry} shows Berry curvature plot of a Weyl semimetal in the $k_z=0$ plane, where we assumed the node separation is $\mathbf{k}_0=(0.5,0.5,0)$. At the origin is a Weyl node with chirality quantum number +1 which acts as a source of Berry flux. At $\mathbf{k}_0$ we have another node with chirality quantum number -1 which acts a sink of Berry flux.

Let us first concern ourselves with the normal contribution to the Nernst effect i.e. due to an external magnetic field. As we pointed out earlier, this contribution is further modified due to effects of the Berry curvature, which are encoded in the factors $c_x$, $c_y$ and $D(\mathbf{B},\Omega_{\mathbf{k}})$, expressed in Eq.~\ref{axx2}-~\ref{syx2}. Adding up the contribution from both the Weyl nodes and keeping terms only upto linear order in the magnetic field $\mathbf{B}=B\hat{z}$ (which is justified in the limit $\omega\tau\ll 1$), the zero temperature longitudinal electrical conductivity for the Weyl system becomes
\begin{align}
\sigma_{xx} &= 2e^2 \int {[d\mathbf{k}]} \frac{v_x^2\tau}{1-(eB\Omega_z/\hbar)^2} \delta(\epsilon_{\mathbf{k}}-\mu)\nonumber\\
&=\frac{e^2\tau\mu^6}{4\pi^2\hbar^3 v_F}\int_0^{\pi}{\frac{\sin^3\theta d\theta}{\mu^4-\hbar^2 e^2 v_F^4 B^2 \cos^2\theta}}
\label{sxx_weyl}
\end{align}
For $B\sim$ 1T, $v_F\sim 10^5 m/s$, (which is the typical value for $v_F$ in a WSM\cite{Wan:2011}), the factor $\hbar^2 e^2 v_F^4 B^2\sim10^{-10}$. Thus away from $\mu=0$, the expression reduces to Eq.~\ref{sxx_dirac} for a Dirac node, except upto an overall multiplicative factor of 2 (for two nodes). At $\mu=0$, where the Fermi surface just reduces to a pair of Weyl points, $\sigma_{xx}=0$. From Eq.~\ref{syx2}, we can calculate the charge Hall conductivity $\sigma_{xy}$ for a WSM (again upto linear order in B):
\begin{align}
\sigma_{yx} &= 2e^2 \int {[d\mathbf{k}]} \frac{v_y^2\omega\tau^2 (1+(eB\Omega_z/\hbar)^2)}{(1-(eB\Omega_z/\hbar)^2)^2} \delta(\epsilon_{\mathbf{k}}-\mu)\nonumber\\
&=\frac{e^2\omega\tau^2}{4\pi^2\hbar^3v_F}\int_0^{\pi}{\frac{\sin^3\theta\mu^6 (\mu^4 + \hbar^2 e^2 v_F^4 B^2 \cos^2\theta)}{(\mu^4 - \hbar^2 e^2 v_F^4 B^2 \cos^2\theta)^2}d\theta}
\label{sxy_weyl}
\end{align}
Further away from $\mu=0$, the above expression also reduces to Eq.~\ref{sxy_dirac} for  Dirac node, after making the substitution for the cyclotron frequency $\omega=ev_F^2 B/\hbar \mu$. At $\mu=0$, again $\sigma_{xy}=0$. We can therefore conclude that away from the band touching point (which line-up with the fine tuning of chemical potential $\mu=0$), and for typical values of the Fermi velocity and weak magnetic field (such that the semiclassical Boltzmann approach is still valid) the deviation of the normal Nernst response due to the Berry curvature is negligible. In this limit the normal contribution to the Nernst response roughly reduces to the sum of individual contributions from the two Dirac nodes. We have also verified this conclusion through explicit numerical integration, even at finite temperatures, using Eq.~\ref{axx2}-~\ref{syx2}. 

\begin{figure}[h]
\includegraphics[scale=.21]{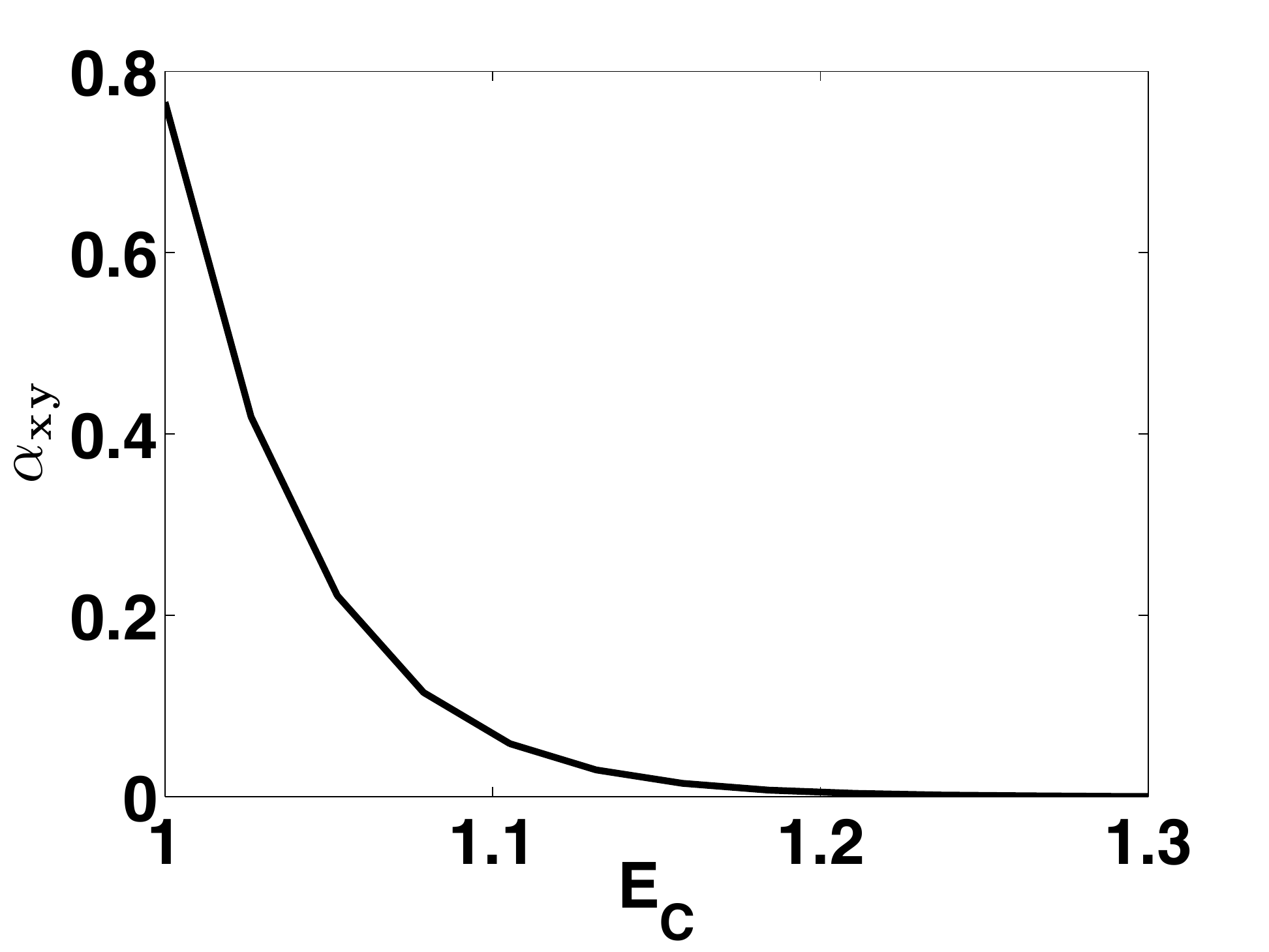}
\includegraphics[scale=.21]{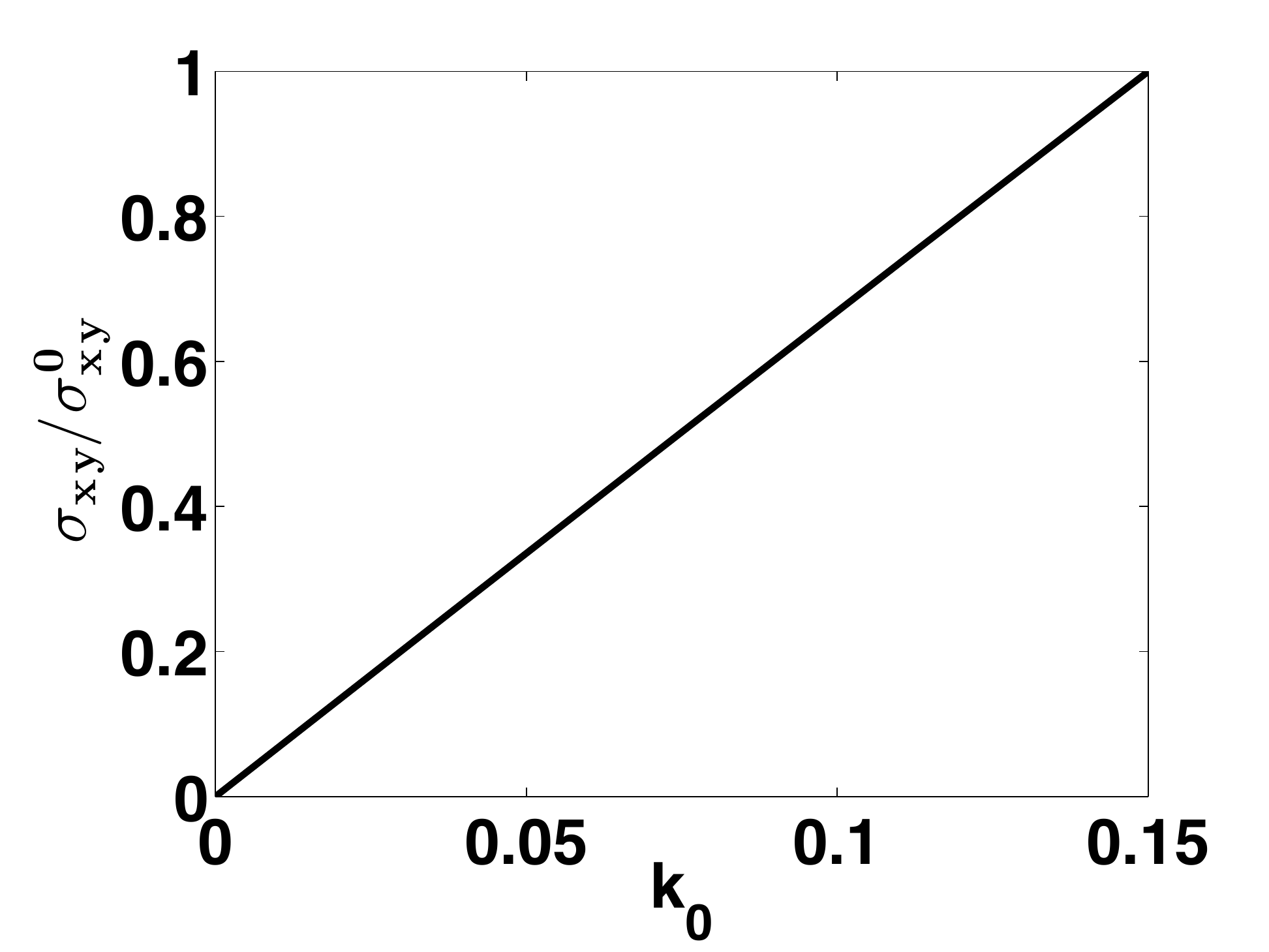}
\caption{\textit{Left}: Plot of anomalous $\alpha_{xy}$ (in the units of $\mu k_Be/\hbar$) for a bounded linear WSM with two bounded linear Dirac nodes vs. the upper energy cutoff $E_C$ (in the units of 10meV), for $\mu=-0.7t$ and $k_0=0.25$. As the upper energy cutoff for each bounded Dirac node increases, $\alpha_{xy}$ decreases eventually becoming close to zero. \textit{Right}: Anomalous Hall conductivity $\sigma_{xy}$ as a function of the node separation $k_0$ for a bounded linear WSM, where $\sigma^0_{xy}$ is the Hall conductivity when $k_0=0.15$.}
\label{anom_weyl_plot1}
\end{figure}

A Weyl system also exhibits anomalous Nernst response even at zero-field and thus the total Nernst signal must arise from both contributions. The anomalous Hall conductivity $\sigma^A_{xy}$ for a time-reversal broken Weyl semimetal is non-zero and varies linearly with the node separation ${k}_0$,\cite{Burkov:2011} which can be obtained by integrating Eq.~\ref{syx2} with the correct regularization that is consistent with the broken symmetries in the presence of $k_0$.~\cite{Goswami:2013}
\begin{eqnarray}
\sigma^A_{xy} = -\frac{e^2}{\hbar}\frac{{k}_0}{2\pi^2}
\label{sxy_weyl1}
\end{eqnarray}
The result in Eq.~\ref{sxy_weyl1} suggests that the $\sigma^A_{xy}$ remains unaltered with temperature or for a finite $\mu$. This result is strictly valid only for an unbounded linear dispersion of Dirac fermions. Mott formula (Eq.~\ref{Mott_eqn}) then suggests that $\alpha^A_{xy}=0$. If an upper physical cutoff on the energy of a Dirac node is imposed then $\alpha^A_{xy}$ is non-zero, because the contributions from the partially filled states generically will remain finite. Figure~\ref{anom_weyl_plot1} shows the plot for $\sigma^A_{xy}$ as function of node separation $k_0$ for a linear WSM with an upper energy cut-off, and also the plot for $\alpha^A_{xy}$ as a function of upper energy cutoff, obtained using Eq.~\ref{ayx2} and ~\ref{syx2}. Lowering the cutoff results in a finite non-zero $\alpha_{xy}$, thus the anomalous Nernst response is also expected to be non-zero. 

It is not entirely evident from Eq.~\ref{nernst_eqn}, that the anomalous Nernst response will vanish for a linearized Weyl Hamiltonian with an unbounded dispersion when $\alpha^A_{xy}=0$, because of the non-zero factor $\alpha_{xx}\sigma^A_{xy}$ in the numerator of Eq.~\ref{nernst_eqn}. However, using the results from the previous subsection we have from Eq.~\ref{sxx_dirac_2} that near $\mu\rightarrow 0$, $\sigma_{xx}=\sigma^0_{xx}\mu^4$, and from Mott relation: $\alpha_{xx}=\alpha^0_{xx}\mu^3$, thus in the vicinity of the Dirac point, the anomalous Nernst coefficient becomes:
\begin{align}
\vartheta = \frac{\sigma^A_{xy}\alpha^0_{xx}\mu^3}{(\sigma^0_{xx}\mu^4)^2 + (\sigma^A_{xy})^2}
\end{align}
Two points can be noted from the above equation: the anomalous Nernst coefficient vanishes at the Dirac point when $\mu=0$, and the Nernst coefficient is an odd function of $\mu$. The evaluation of total Nernst signal, will have contributions from both normal and anomalous Hall conductivities, and therefore an asymmetric behavior of the total Nernst coefficient about $\mu=0$ is expected.
\section{Nernst effect in a lattice Weyl Hamiltonian}
The novel semimetallic state with Weyl-like fermionic excitations has been recently realized in a series of experiments. In the inversion asymmetric crystalline compound TaAs, experiments have claimed to host topologically non-trivial Fermi arcs and Weyl cones~\cite{Su:2015, Huang:2015, Lv:2015}. Another pathway of observing Weyl fermions in condensed matter is to break TR symmetry by applying a magnetic field in a 3D Dirac semimetal and split each Dirac cone into a pair of Weyl nodes. 3D materials Na$_3$Bi, Cd$_3$As$_2$, and Bi$_{1-x}$Sb$_x$ with $x\sim 3-4\%$ were recently proposed~\cite{Wang:2012, Wang:2013, Fu:2007, Teo:2008, Guo:2011}, and also realized experimentally to be Dirac semimetals~\cite{Liu:2014, SyXu:2013, Neupane:2014, ZKLiu:2014, Borisenko:2014, Jeon:2014, TLiang:2014, Xiong1:2015, Xiong:2015, Kim:2013}. This has paved the way to realizing a TR broken WSM phase, which has been so far verified by a few experimental signatures~\cite{Huang:2015, Xiong:2015, Li:2014, Kim:2013}.
 
One can construct an effective Hamiltonian $H_D(\mathbf{k})$ describing electron dynamics near the symmetry points in momentum space of a 3D Dirac semimetal
\begin{align}
H_D(\mathbf{k}) = v_F (\mathbf{k}\cdot\sigma) \tau_z + m(\mathbf{k})\tau_x,
\label{H_D_eqn}
\end{align}
where $\sigma$ and $\tau$ are the vectors of Pauli spin matrices acting on the spin and pseudo-spin degrees of freedom respectively. The mass term $m(\mathbf{k})$ can be tuned to vanish at isolated points in the Brillouin zone  by choosing a specific form of $m(\mathbf{k})$ as a function of crystal momentum $\mathbf{k}$.  The mass term $m(\mathbf{k})$ can be chosen to be $m(\mathbf{k})=m+\rho\cos(\mathbf{k})$, for $|\mathbf{k}|\ll 1$,  which can be realized physically. For instance, at $x\sim 3-4\%$ in Bi$_{1-x}$Sb$_x$ the mass term can be tuned to zero at particular $\mathbf{K}$ values. This yields a degenerate linearly dispersing 4-component Dirac fermion described by Eq.~\ref{H_D_eqn} with $m(\mathbf{K})=0$. A perturbative Zeeman field coupled with the spin degree of freedom described by $H_Z=gB_z\sigma_z$, can now lift the degeneracy of the Dirac fermion, with the energy spectrum given by~\cite{Kim3:2014}
\begin{align}
E(\mathbf{k}) = \pm\sqrt{v_F^2(k_x^2+k_y^2) + (gB_z\pm v_F k_z)^2},
\label{E_k_dirac_eqn}
\end{align}
where the band-touching point is shifted from $\Gamma$ point at $\mathbf{K}=(0,0,0)$ to $\pm \mathbf{K}_0=\pm(0,0,gB/v_F)$, where $g$ is the Land\'{e} g-factor. The direction along which the two nodes appear is along the direction of the applied magnetic field. The separation $k_0$ between the two Weyl nodes is magnetic field dependent and is given by $k_0=2gB/v_F$. Now the Hamiltonian around each $\mathbf{K}_0$ is that of a linearized Weyl fermion. 

\begin{figure}[h]
\includegraphics[scale=.21]{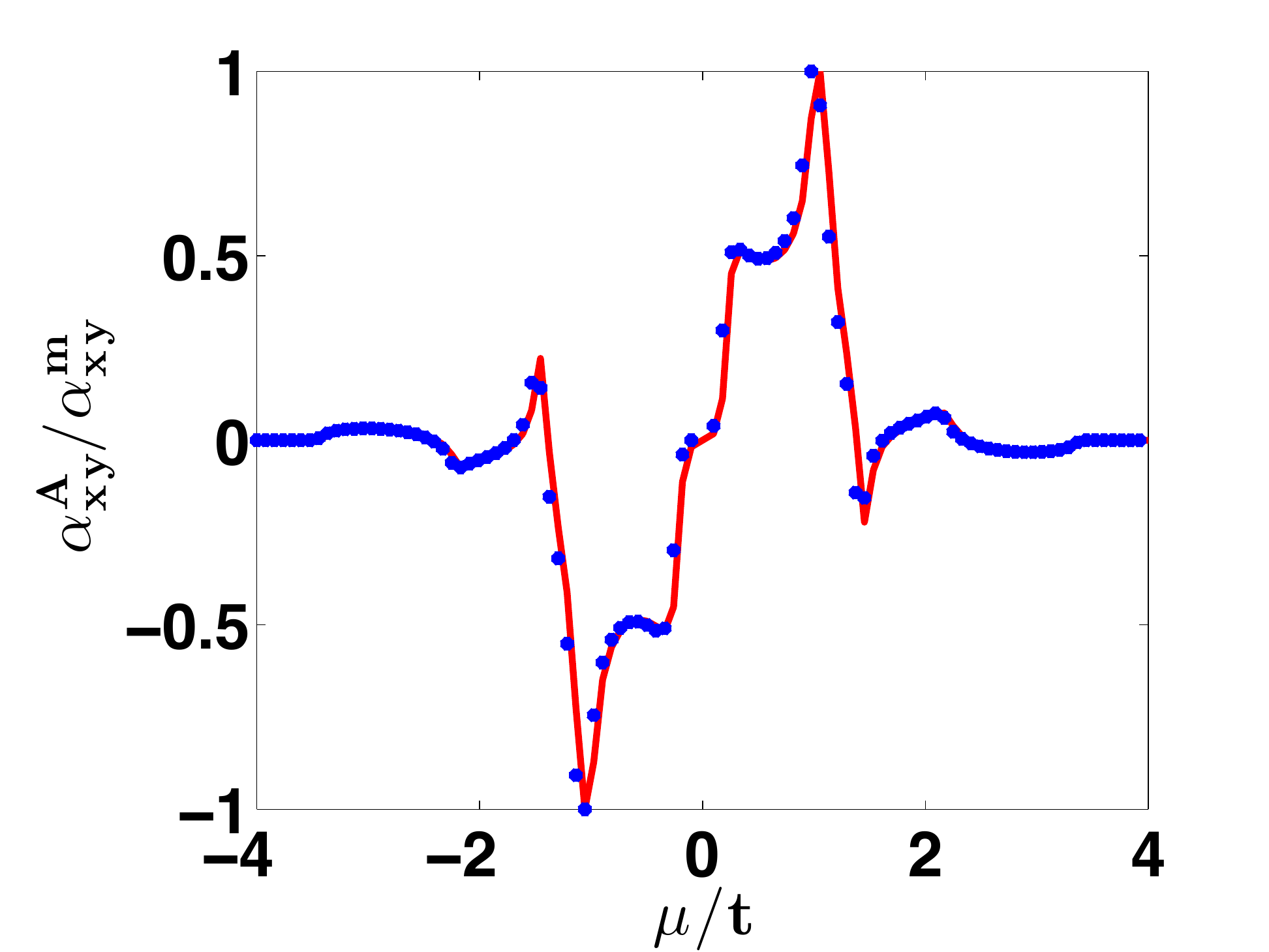}
\includegraphics[scale=.21]{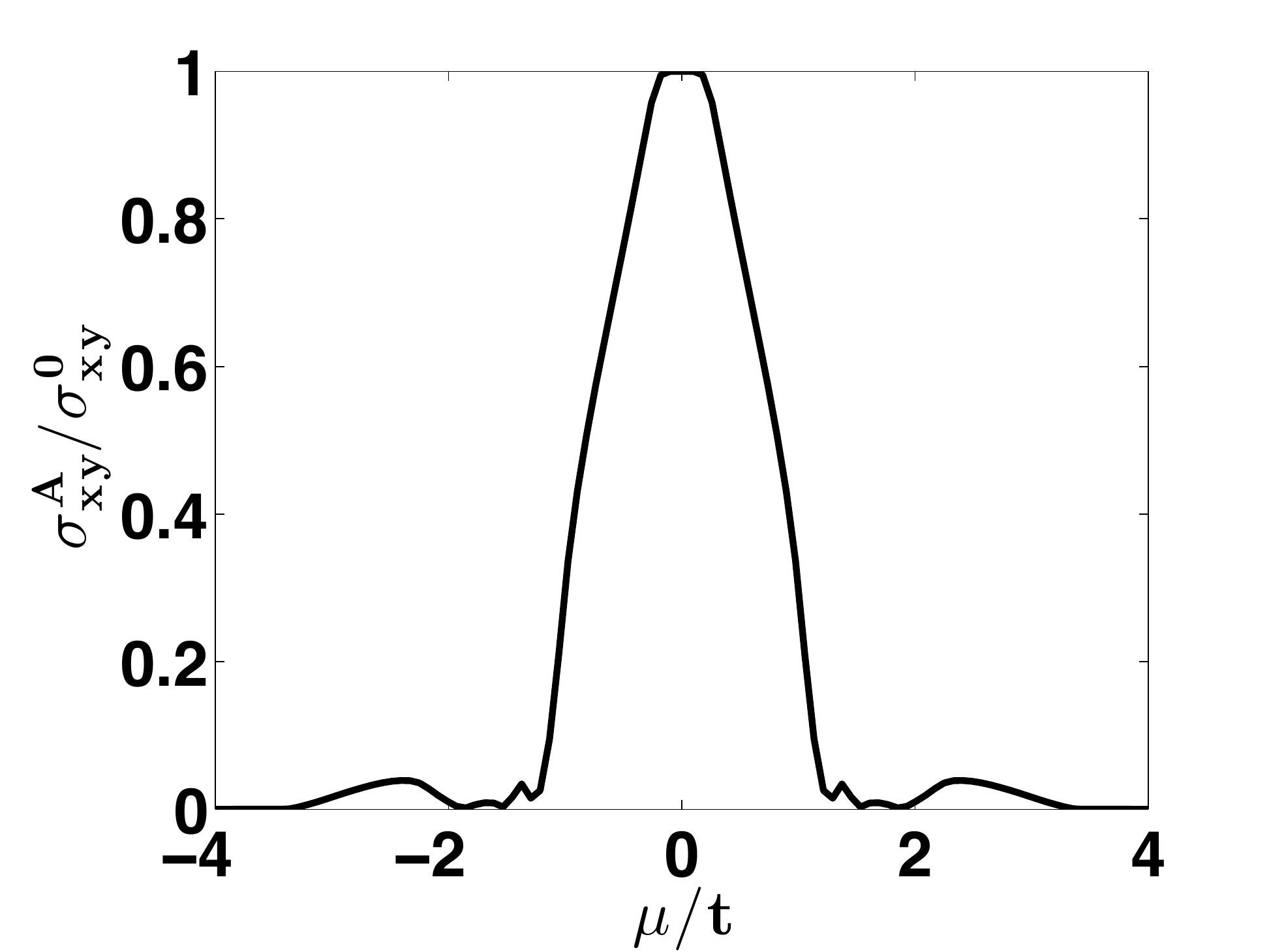}\\
\includegraphics[scale=.21]{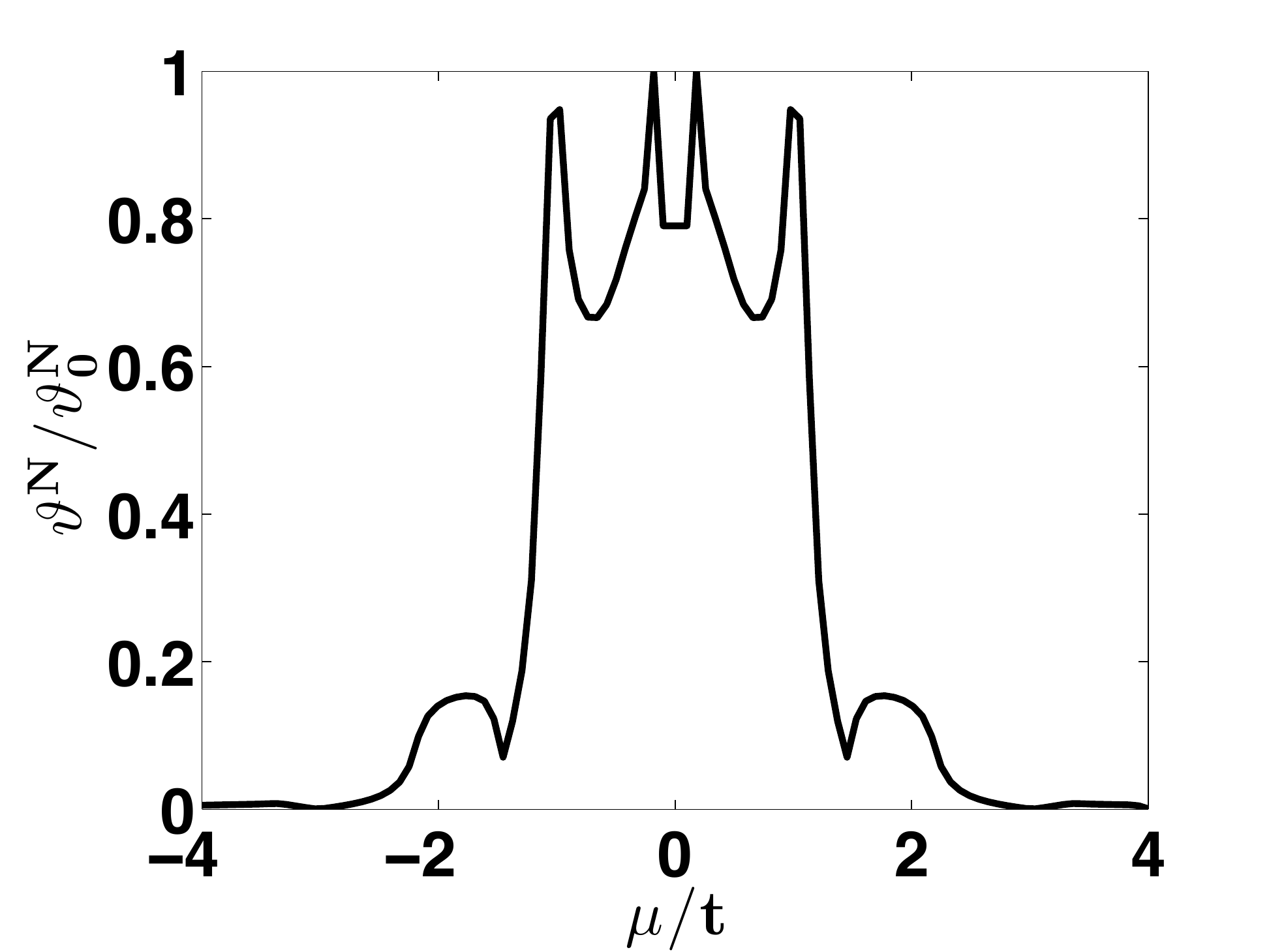}
\includegraphics[scale=.21]{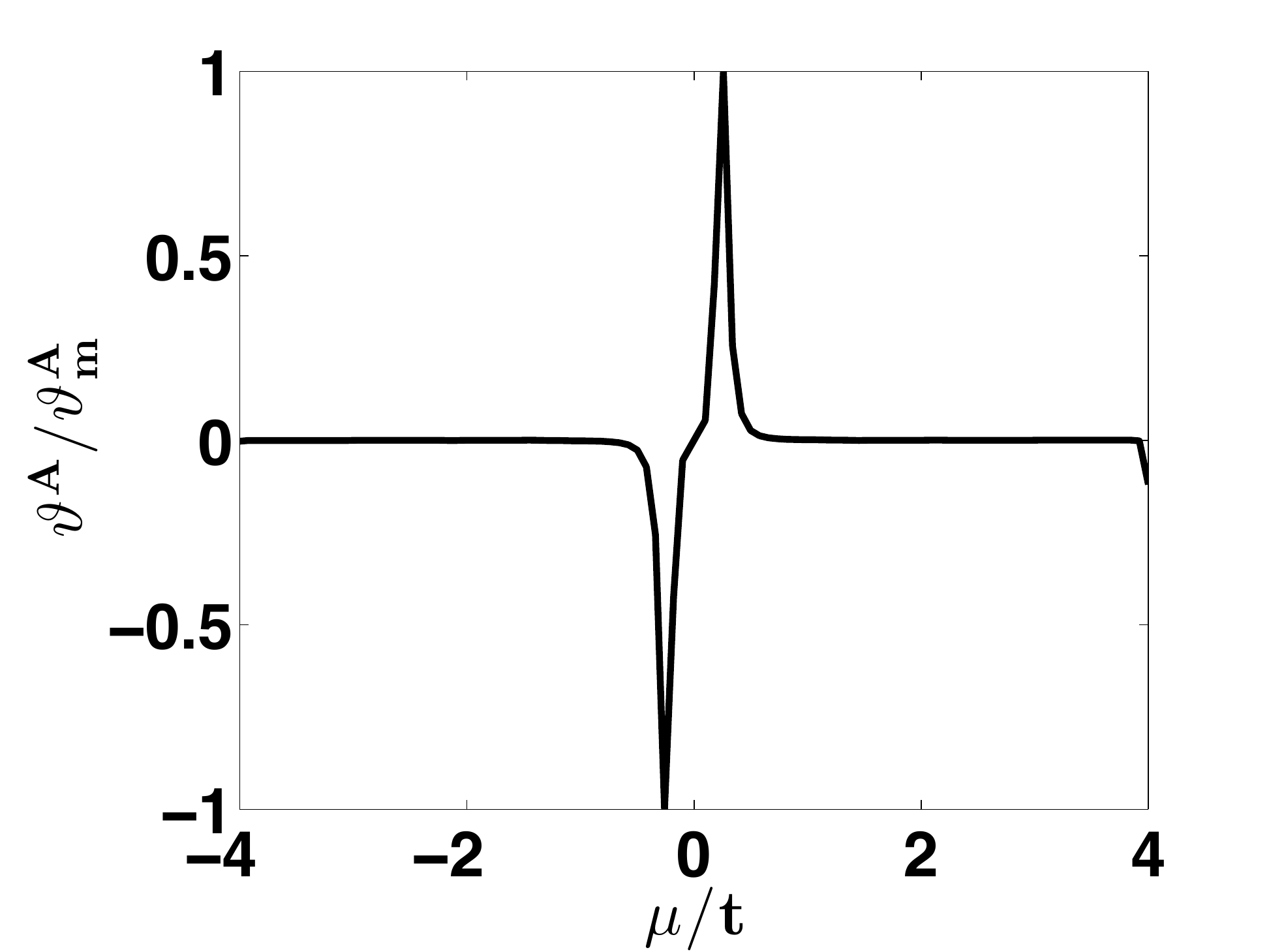}\\
\includegraphics[scale=.21]{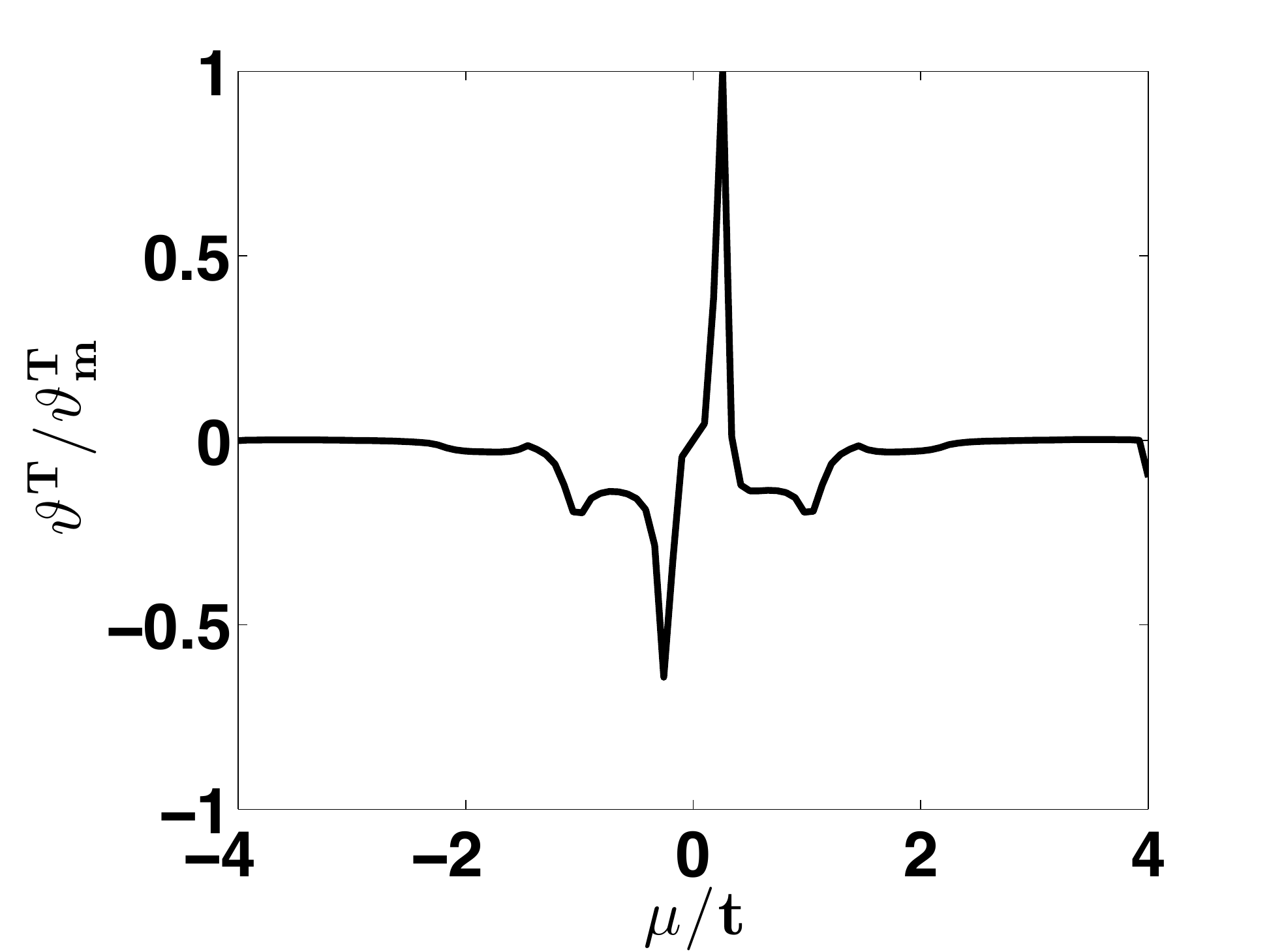}
\includegraphics[scale=.21]{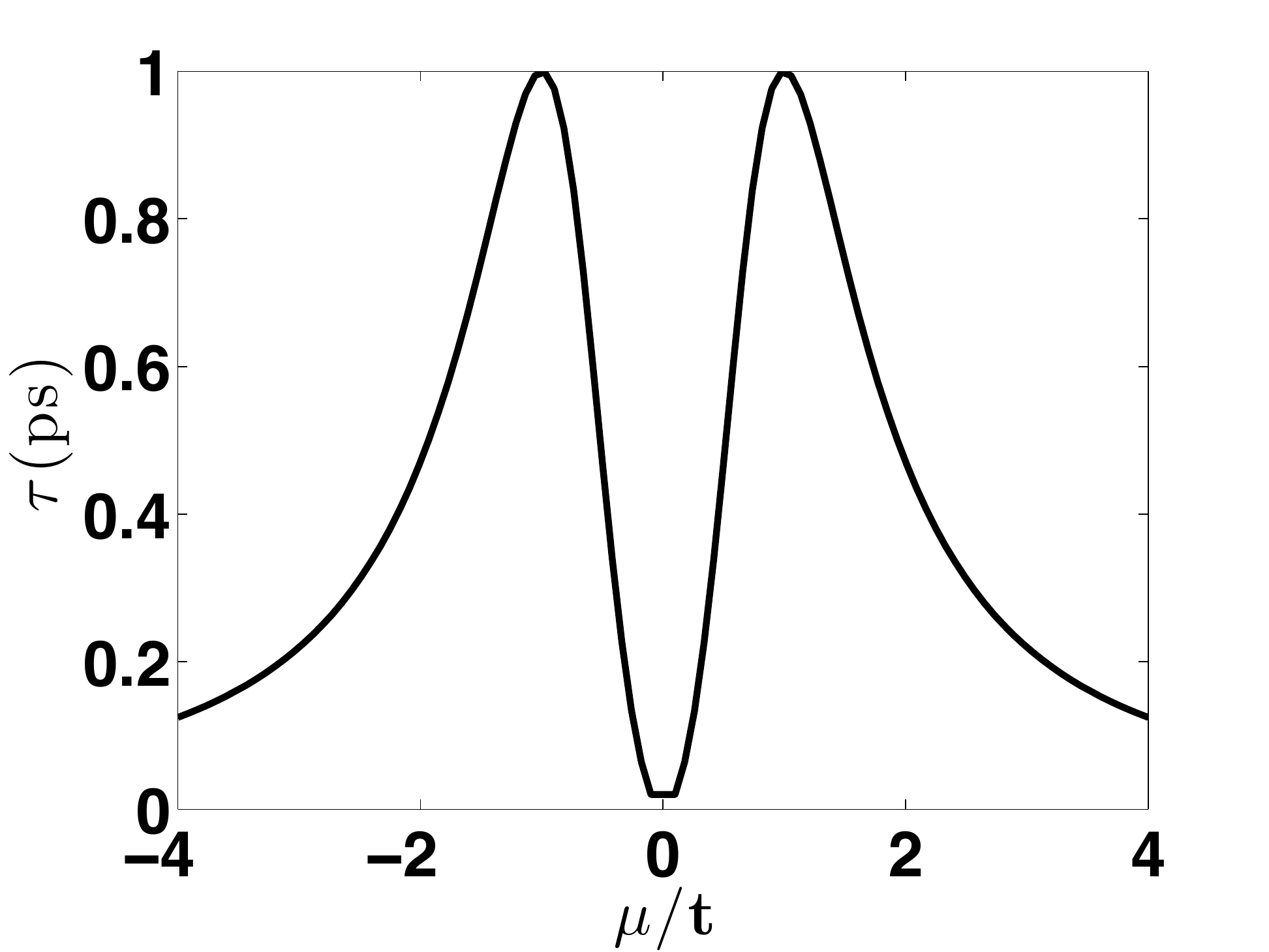}
\caption{(color online) Results for the lattice WSM described by Eq.~\ref{lattice_eqn}. \textit{Top Panel (left)}: In red is the anomalous contribution to $\alpha_{xy}/\alpha^m_{xy}$ obtained using Eq.~\ref{ayx1} , and in blue is the anomalous contribution to $\alpha_{xy}$ using Mott relation (Eq.~\ref{Mott_eqn}) displaying a reasonable agreement; $\alpha^m_{x}$ is the value at $\mu=t$. \textit{Top panel (right)}: Anomalous Hall conductivity as a function of $\mu/t$ (again $\sigma^0_{xy}$ is the value at $\mu=0)$) obtained using Eq.~\ref{syx1}. \textit{Middle panel (left)}: Normal Nernst response $\vartheta^N/\vartheta_0$ as a function of $\mu$, where $\vartheta_0$ is the Nernst coefficient at $\mu=0$. \textit{Middle panel (right)}: Anomalous Nernst response $\vartheta^A/\vartheta_m$ as a function of $\mu$, where $\vartheta_m$ is the anomalous Nernst coefficient at $\mu=0.3t$. \textit{Bottom panel (left)}: Total Nernst coefficient $\vartheta^T/\vartheta_m$ including both normal and anomalous response as a function of $\mu/t$, where $\vartheta_m$ is the Nernst coefficient at $\mu=0.1t$. A slight asymmetry for the total Nernst signal about $\mu=0$ can be noted. \textit{Bottom panel (right)}: Chosen scattering time in $ps$ as a function of $\mu$, as given by Eq.~\ref{total_tau_eqn_1}. The chosen parameters values for this calculation are: $T=20K$, $t=0.1$ $eV$, $m=0.6t$.}
\label{lattice_bands}
\end{figure}

To discuss the Nernst response in a physical Weyl system, it is advantageous to consider a lattice model of Weyl fermions with the lattice regularization providing a physical ultra-violet smooth cut-off to the low energy Dirac spectrum, because the linearized continuum theory for calculating the anomalous Hall current at a finite density turns out to be insufficient. For simplicity, instead of considering the band Hamiltonian for Bi$_{1-x}$Sb$_x$, we consider a prototype lattice Hamiltonian for Weyl fermions described by $H_{latt}(\mathbf{k})$ in Eq.~\ref{lattice_eqn}, because it serves our purpose of discussing the Nernst effect.
\begin{align}
H_{latt}(\mathbf{k}) &= t(\sin (k_xa) \sigma_x + \sin (k_yb) \sigma_y + \cos (k_zc) \sigma_z) \nonumber \\
 &+ m(2-\cos (k_xa)-\cos (k_yb))\sigma_z\equiv \mathbf{N}_{\mathbf{k}}\cdot\boldsymbol\sigma,
 \label{lattice_eqn}
\end{align}
$H_{latt}(\mathbf{k})$ supports a pair of Weyl fermions located at $(0,0,\pm \pi/2c)$,thus $\mathbf{k}_0=(0,0,\pi/c)$ when $m>t/2$. This lattice model of Weyl fermions described in Eq.~\ref{lattice_eqn} can mimic the experimentally relevant TR breaking WSM model described in Eq.~\ref{H_D_eqn} and the paragraph below it, when we consider an external magnetic field $\mathbf{B}=(0,0,\pi v_F/g)$, as the node separation is fixed from Eq.~\ref{lattice_eqn}. This is however sufficient to discuss Nernst effect in the lattice WSM. The energy-band spectrum is shown in Figure~\ref{weyl_2schematic} obtained from diagonalising the Hamiltonian in Eq.~\ref{lattice_eqn}. The anomalous Hall conductivity $\sigma^A_{xy}$ for this Hamiltonian is given by Eq.~\ref{syx1} at finite temperature and chemical potential, which is plotted in Figure~\ref{lattice_bands}. At zero temperature and at $\mu=0$, $\sigma^A_{xy}=e^2/h$. The $i^{{th}}$ component of Berry curvature vector $\mathbf{\Omega}_{\mathbf{k}}$, for $H_{latt}(\mathbf{k})$ is given by
\begin{align}
\Omega_{\mathbf{k},n,i} = (-1)^n\epsilon_{ijl}\frac{\mathbf{N}_{\mathbf{k}}\cdot\[\frac{\partial \mathbf{N}_{\mathbf{k}}}{\partial k_j}\times\frac{\partial \mathbf{N}_{\mathbf{k}}}{\partial k_l}\]}{4|\mathbf{N}_{\mathbf{k}}|^3}
\label{berry_band_eqn}
\end{align}
In Eq.~\ref{berry_band_eqn}, $n$ stands for the band index. The anomalous Peltier coefficient $\alpha^A_{xy}$, can be calculated using Eq.~\ref{ayx1} and is non-zero, as shown in Figure~\ref{lattice_bands}, because $\alpha^A_{xy}$ is no longer a constant function of $\mu$. Figure~\ref{lattice_bands} also shows $\alpha^A_{xy}$ calculated using the Mott relation. This feature of non-zero $\alpha^A_{xy}$ was absent in the linearized model of a WSM. Figure~\ref{lattice_bands} shows the total Nernst response $\vartheta$ obtained by numerical calculation using Eq.~\ref{nernst_eqn},~\ref{axx2}-\ref{syx2} for this lattice model. We observe that the normal Nernst coefficient is non-zero around $\mu=0$, and is an even function of $\mu$, which consistent with the findings of Sec. III.  In Sec. III, we also pointed out that the anomalous Nernst coefficient vanishes at $\mu=0$ and is an odd function of $\mu$, and thus the total Nernst response is expected to show an asymmetric behavior about $\mu=0$. These conclusions do no change for a lattice model, and a slight asymmetry can be observed in the plot of Nernst coefficient in Figure~\ref{lattice_bands}.

\section{Magneto-thermal conductivity for a Weyl semimetal}
When a temperature gradient $\nabla T$ is applied across a sample, a charge current $\mathbf{J}$ is developed. In the absence of a charge current, from Eq.~\ref{thermal}, we must have $\mathbf{E} = \hat{\sigma}^{-1}\hat{\alpha}\nabla T$. Using this expression for $\mathbf{E}$ in the formula for the thermal current $\mathbf{Q}$ given in Eq.~\ref{thermal}, we can write the following linear response relation
\begin{align}
\mathbf{Q} = (T\hat{\alpha}\hat{\sigma}^{-1}\hat{\alpha} - \hat{l})\nabla T = \hat{\kappa} (-\nabla T)
\end{align}
$\hat{\kappa}$ is the thermal conductivity tensor whose longitudinal and Hall components can be written explicitly as:
\begin{align}
\kappa_{xx} = l_{xx} - \frac{\sigma_{xx}(\alpha_{xx}^2 - \alpha_{xy}^2) + 2\sigma_{xy}\alpha_{xx}\alpha_{xy}}{\sigma_{xx}^2 -\sigma_{xy}^2}
\label{kxx_1_eqn}
\end{align}
\begin{align}
\kappa_{xy} = l_{xy} - \frac{\sigma_{xy}(\alpha_{xy}^2 - \alpha_{xx}^2) + 2\sigma_{xx}\alpha_{xx}\alpha_{xy}}{\sigma_{xx}^2 -\sigma_{xy}^2}
\label{kxy_1_eqn}
\end{align}
Usually $\hat{l}$ is identified with $\hat{\kappa}$, however the second term in Eq.~\ref{kxx_1_eqn} and Eq.~\ref{kxy_1_eqn} is still non-zero, though small compared to $\hat{l}$. Wiedemann-Franz law states that the ratio of thermal conductivity $\kappa$ and electrical conductivity $\sigma$ for a metallic state is proportional to temperature. The law holds for a generic system as long as it can be termed as Landau Fermi liquid where the quasiparticle description of electronic states remains valid, 
\begin{align}
\frac{\kappa_{ij}}{\sigma_{ij}} = L_0 T
\label{wf_law_eqn}
\end{align}
Eq.~\ref{wf_law_eqn} states the Wiedemann-Franz law, where $L_0$ is the Lorenz number ($L_0=\pi^2k_B^2/3e^2$). 

\begin{figure}[h]
\includegraphics[scale=.21]{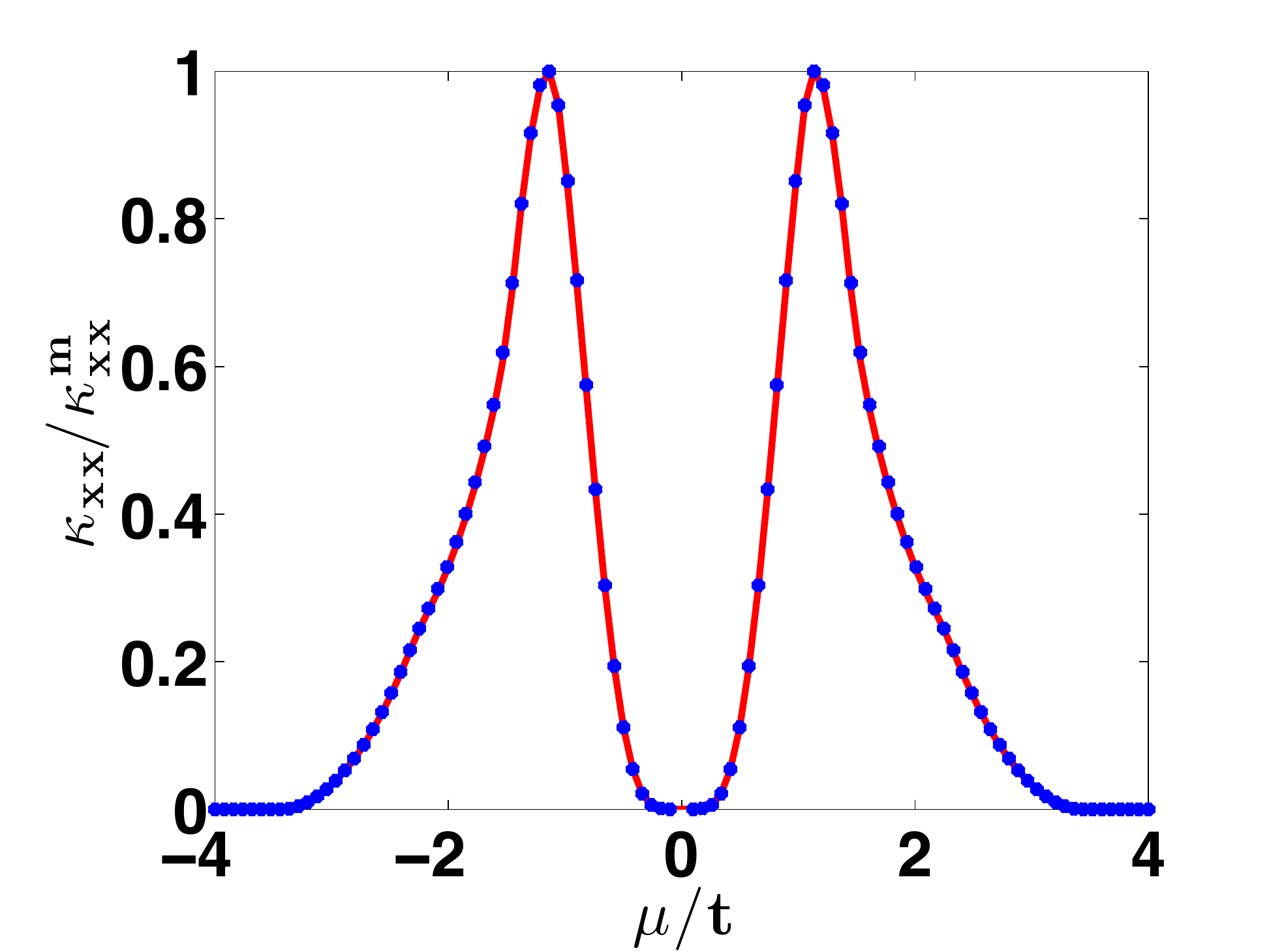}
\includegraphics[scale=.21]{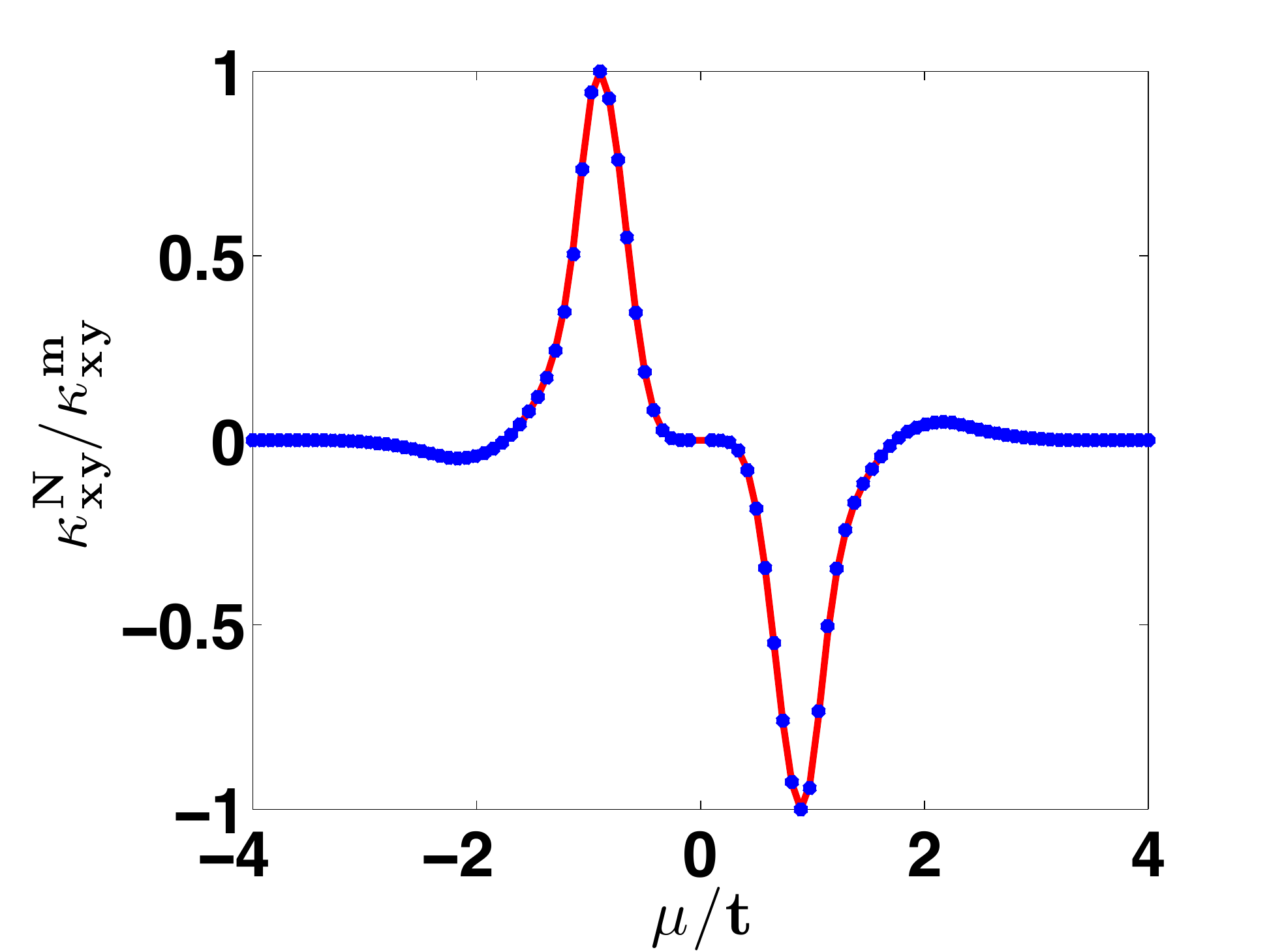}\\
\includegraphics[scale=.2]{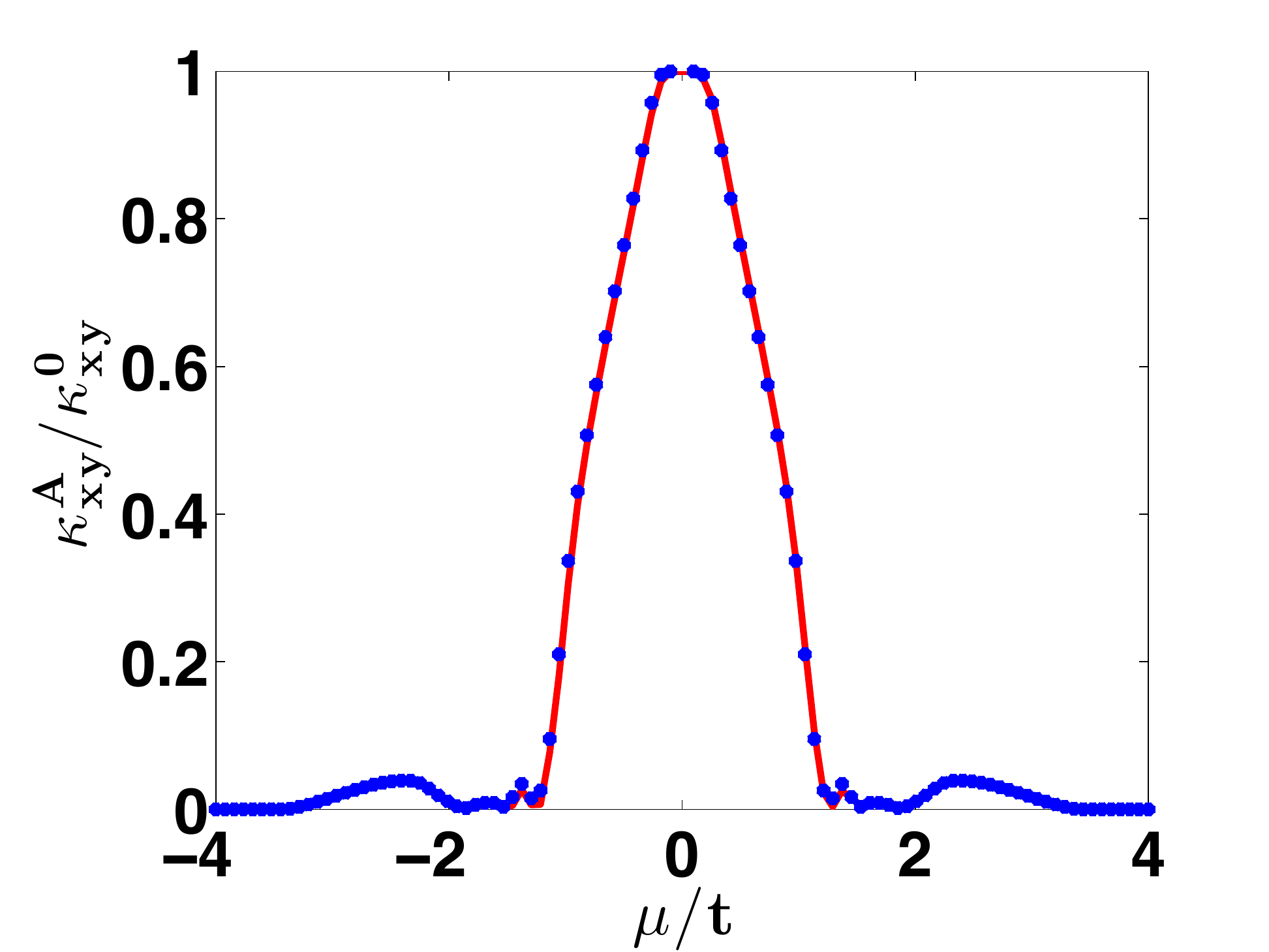}
\includegraphics[scale=.21]{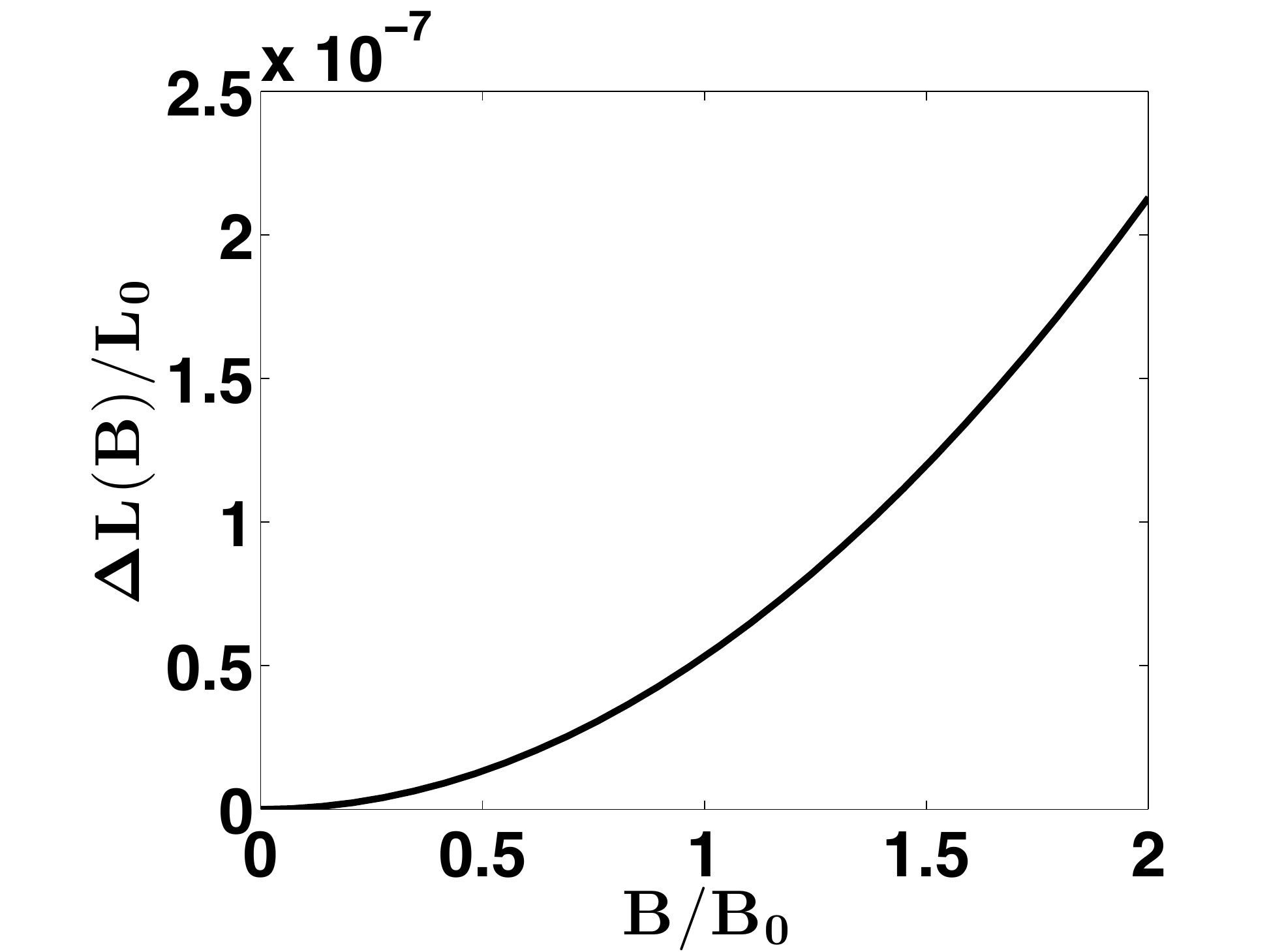}
\caption{(color online) \textit{Top panel (left)}: Longitudinal magneto-thermal conductivity $\kappa_{xx}$ for the WSM lattice model as a function of $\mu/t$, where $\kappa^m_{xx}$ is the value at $\mu=t$. \textit{Top panel (right)}: Normal $\mathbf{B}$-dependent contribution to $\kappa^N_{xy}$, where $\kappa^m_{xy}$ is the value at $\mu=t$. \textit{Bottom panel (left)}: Anomalous zero magnetic-field contribution to $\kappa^A_{xy}$, where $\kappa^0_{xy}$ is the value at $\mu=0$. In all the three figures, we have plotted in red the thermal conductivity obtained using Eq.~\ref{lxx2} and Eq.~\ref{lyx2}, and in blue using Wiedemann-Franz law Eq.~\ref{wf_law_eqn} (all three plots are for transverse setup i.e. $\nabla T\perp\mathbf{B}$). \textit{Bottom panel (right)}: Plot of $\Delta L(B)/L_0=(L(B)/L_0-1)$ as a function of applied magnetic field $B$ (here $B_0=v_Fk_0/2g$) when $\nabla T\parallel\mathbf{B}$ showing an additional $B^2$ dependence of the Lorenz number arising from the chiral anomaly term $(\mathbf{\Omega}\cdot\mathbf{v})$.}
\label{kappa_plots}
\end{figure}

In the absence of a magnetic field, the $\mathbf{B}$-dependent contribution to $\sigma_{xy}$ is zero, and $\kappa_{xx}=l_{xx}-\alpha_{xx}^2/\sigma_{xx}$ will be given by the standard expression of the longitudinal thermal conductivity: 
\begin{align}
l_{xx} = \int{[d\mathbf{k}] v_x^2\left(\tau\frac{(\epsilon-\mu)^2}{T} \left(-\frac{\partial f_{eq}}{\partial\epsilon}\right) \right)} 
\label{lxx_1_eqn}
\end{align}
To discuss magneto-thermal conductivity, we first examine the case which is also relevant to the Nernst experimental setup discussed in Section II, III and IV i.e. $\nabla T=(\partial T/\partial x) \hat{x}$ and $\mathbf{B}=B\hat{z}$. The expressions for charge and thermoelectric conductivities $\hat{\sigma}$ and $\hat{\alpha}$ have been already obtained in Eq.~\ref{axx2}-\ref{syx2}.  From Eq.~\ref{Q1_eqn} and Eq.~\ref{thermal}, we can read the conductivity tensor $\hat{l}$ as
\begin{align}
l_{xx} = \int{[d\mathbf{k}] v_x^2\left(\tau\frac{(\epsilon-\mu)^2}{T} \left(-\frac{\partial f_{eq}}{\partial\epsilon}\right) \left(c_x-D \right) \right)}
\label{lxx2}
\end{align}
\begin{align}
&l_{yx} = \int{[d\mathbf{k}] (v_y^2 c_y+v_xv_y (c_x-D))\left(\tau\frac{(\epsilon-\mu)^2}{T} \left(-\frac{\partial f_{eq}}{\partial\epsilon}\right) \right)} \nonumber \\ &  + \frac{k_B\nabla T}{\beta\hbar}\times\int{[d\mathbf{k}] \mathbf{\Omega}_{\mathbf{k}}\(\frac{\pi^2}{3}f_{eq}+\beta^2(\epsilon-\mu)^2f_{eq}\)}\nonumber \\
& - \frac{k_B\nabla T}{\beta\hbar}\times\int{[d\mathbf{k}] \mathbf{\Omega}_{\mathbf{k}}(\mbox{ln}(1+e^{-\beta(\epsilon_{\mathbf{k}}-\mu)})^2+2\mbox{Li}_2(1-f_{eq}))}
\label{lyx2}
\end{align}
The $\mathbf{B}$-dependent longitudinal magneto-thermal conductivity ${l}_{xx}$ is further modified from Eq.~\ref{lxx_1_eqn} by the factor of $(c_x-D)$ which is a function of the Berry curvature. The first term of the transverse magneto-thermal conductivity $l_{yx}$ in Eq.~\ref{lyx2} is the standard $\mathbf{B}$-dependent contribution. The second and the third terms give the zero magnetic-field anomalous thermal conductivity. The Wiedemann-Franz law given in Eq.~\ref{wf_law_eqn} remains valid for $\kappa_{xx}$ and $\kappa_{xy}$ as shown in Figure~\ref{kappa_plots}.  

A more interesting scenario occurs when $\nabla T=(\partial T/\partial z) \hat{z}$, $\mathbf{E}=0$, $\mathbf{B}=B\hat{z}$ i.e when the applied temperature gradient is parallel to the magnetic field. 
Using Eq.~\ref{rdot_eqn} and Eq.~\ref{pdot_eqn}, the steady state Boltzmann equation (Eq.~\ref{boltz_basic_eqn}) becomes
\begin{align}
&\(\frac{\epsilon-\mu}{T}\nabla_z T \(-\frac{\partial f_{eq}}{\partial \epsilon}\)\)\(v_z + \frac{e}{\hbar}(\mathbf{v}\cdot\mathbf{\Omega}_{\mathbf{k}})\) \nonumber \\
&+\frac{eB}{\hbar^2}\(v_y\frac{\partial}{\partial k_x} - v_x \frac{\partial}{\partial k_y}\)f_{\mathbf{k}} = -\frac{f_{\mathbf{k}}-f_{eq}}{D\tau}
\label{boltz_3_eqn}
\end{align}
The following Ansatz is chosen for the distribution function $f_{\mathbf{k}}$, which is a solution of Eq.~\ref{boltz_basic_eqn},
\begin{align}
f_{\mathbf{k}}-f_{eq} &= -D\tau\frac{\epsilon-\mu}{T}\nabla_z T \(-\frac{\partial f_{eq}}{\partial\epsilon}\)\(v_z + \frac{e}{\hbar}B(\mathbf{v}\cdot\mathbf{\Omega}_{\mathbf{k}})\)\nonumber \\
& +\(-\frac{\partial f_{eq}}{\partial\epsilon}\)\mathbf{v}\cdot\mathbf{\Lambda}
\label{ansatz_3_eqn}
\end{align}
The correction factor $\Lambda$ in the Ansatz for $f_{\mathbf{k}}$ is introduced to account for a perturbative magnetic field $\mathbf{B}$. Substituting for $f_{\mathbf{k}}$ given in Eq.~\ref{ansatz_3_eqn} into the Boltzmann equation (Eq.~\ref{boltz_3_eqn}), and imposing the condition that the equation should remain valid for all values of $\mathbf{v}$, we find that $\Lambda_z=0$. Introducing $V=v_x+iv_y$ and $\Lambda=\Lambda_x-i\Lambda_y$, the Boltzmann equation can be rewritten in the following form
\begin{align}
&\mbox{Re}\(-iV\frac{eBD\tau(\epsilon-\mu)\nabla_z T}{T\hbar^2}\(-\frac{\hbar}{m_{xz}}-er\)\)\nonumber\\
+&\mbox{Re}\(V\frac{eBD\tau(\epsilon-\mu)\nabla_z T}{T\hbar^2}\(\frac{\hbar}{m_{yz}}+es\)\) \nonumber\\
=-&\mbox{Re}\(\frac{-ieBV\Lambda}{m_{xx}\hbar}-\frac{eBV^*\Lambda}{\hbar m_{yx}}+\frac{V\Lambda}{D\tau}\),
\label{boltz_3}
\end{align}
where,
\begin{align}
r=\frac{\Omega_x}{m_{xx}} + \frac{\Omega_y}{m_{xy}} + \frac{\Omega_z}{m_{xz}}\\
s=\frac{\Omega_x}{m_{xy}} + \frac{\Omega_y}{m_{yy}} + \frac{\Omega_z}{m_{yz}}
\end{align}
The factors $\Lambda_x$ and $\Lambda_y$ can be straightforwardly evaluated from the real and imaginary parts of the complex vector $\Lambda=\Lambda_x+i\Lambda_y$, which is a solution of Eq.~\ref{boltz_3}.
Now substituting for the distribution function $f_{\mathbf{k}}$ in Eq.~\ref{J2}, it is then possible to deduce the conductivity tensor $\hat{\alpha}$ and $\hat{l}$. The longitudinal conductivities are obtained to be:~\cite{Fiete:2014, Kim:2014}
\begin{align}
\alpha_{zz} = \frac{e^2}{T}\int{D[d\mathbf{k}] \(v_z + \frac{eB}{\hbar}\mathbf{\Omega}\cdot\mathbf{v}\)^2\tau \({\epsilon-\mu}\)\left(-\frac{\partial f_{eq}}{\partial\epsilon}\right)}
\label{azz}
\end{align}
\begin{align}
l_{zz} = \int{D[d\mathbf{k}] \(v_z + \frac{eB}{\hbar}\mathbf{\Omega}\cdot\mathbf{v}\)^2\tau \(\frac{(\epsilon-\mu)^2}{T}\)\left(-\frac{\partial f_{eq}}{\partial\epsilon}\right)}
\label{lzz}
\end{align}
In a similar fashion one can also calculate the charge conductivity $\sigma_{zz}$.
\begin{align}
\sigma_{zz} = e^2\int{D[d\mathbf{k}] \(v_z + \frac{eB}{\hbar}\mathbf{\Omega}\cdot\mathbf{v}\)^2\tau \left(-\frac{\partial f_{eq}}{\partial\epsilon}\right)}
\label{szz}
\end{align}
The results in Eq.~\ref{azz}-Eq.~\ref{szz} are of interest because of the a $B^2$ dependence arising from the chiral-anomaly term $\mathbf{\Omega}\cdot\mathbf{v}$. Note that this term does not arise in the transverse setup i.e. when $\nabla T$ and $\mathbf{B}$ orthogonal to each other, and is thus linked to the topological $\mathbf{E}\cdot\mathbf{B}$ term arising in axion-electrodynamics of WSM. 
We can write the following simple relations for $\sigma_{zz}$ and $\kappa_{zz}$:
\begin{align}
\sigma_{zz} = \sigma_0 + \alpha B^2\\
\kappa_{zz} = \kappa_0 + \beta B^2,
\end{align}
where $\sigma_0$ and $\kappa_0$ are the longitudinal conductivities for $B=0$. The coefficients $\alpha$ and $\beta$ (not to be confused with $\beta=1/k_BT$) account for the $B^2$ dependence of $\sigma_{zz}$ and $\kappa_{zz}$ respectively. They depend on the band-structure of the Hamiltonain and can be obtained by the $k-$ space integrals defined in Eq.~\ref{szz}-~\ref{lzz}. The Lorenz number $L$ in the Wiedemann-Franz law given in Eq.~\ref{wf_law_eqn}, will be $B$-dependent, and can be written as:
\begin{align}
L(B)=\frac{\kappa_{zz}}{T\sigma_{zz}} = L_0 + \Delta L(B) = \frac{\kappa_0 + \beta B^2}{T(\sigma_0 + \alpha B^2)}\\
\Delta L(B) \approx \frac{(\beta\sigma_0 - \alpha\kappa_0)B^2}{T\sigma_0^2} =  \frac{(\beta\sigma_0 - \alpha L_0T\sigma_0)B^2}{T\sigma_0^2}
\label{Eq_Delta_L_1}
\end{align}
$\Delta L(B)$ gives the $B^2$ enhancement of the Lorenz number from its standard value $L_0$. Figure~\ref{kappa_plots} displays the quadratic behavior of the Lorenz number $L(B)$ with the magnetic field in the parallel setup for the lattice WSM Hamiltonian given in Eq.~\ref{lattice_eqn}, thus showing a violation of the Wiedemann-Franz law. It is also worthwhile to point out the sign of $\Delta L$ in Eq.~\ref{Eq_Delta_L_1}, which will depend on the details of the band structure of the Hamiltonian. The Lorenz number, which is the ratio of thermal to electrical conductivity will increase (decrease) from its standard value if the $B^2$ coefficient in the expression for thermal conductivity is greater (lesser) than electrical conductivity In the present case, the sign of $\Delta L$ was found to be positive. Similar conclusions on the sign of $\Delta L$ were obtained in previous work~\cite{Kim2:2014}.
\section{Conclusions}
In this work we have studied the Nernst response of a time-reversal broken Weyl semimetal. As a consequence of non-zero anomalous Hall response in a Weyl system with broken time-reversal symmetry, it is generally expected that an anomalous Nernst conductivity is also observed. This is because generally the Peltier coefficient which is related to the first derivative of the charge conductivity with respect to the chemical potential should not vanish. However, a linearized Weyl fermionic system was found to have its anomalous Hall conductivity independent of chemical potential and temperature. Previous studies~\cite{Fiete:2014} have therefore argued that the anomalous Peltier coefficient and the anomalous Nernst response for a system of Weyl fermions should be zero. We show this by considering a physical description of a WSM which is cut-off at higher energies by either considering a bounded linearized Weyl Hamiltonian, or a lattice regularization providing a smooth physical ultra-violet cut-off in a lattice model of Weyl fermions. This produces a non zero Peltier coefficient and thus a non-vanishing Nernst response measurable experimentally. 

Starting with the semi-classical Boltzmann approach to linear transport in a system, we first derived the expressions for charge and thermal conductivities in the presence of a perturbative magnetic field and a temperature gradient orthogonal to each other, for a generic band Hamiltonian which has a non-trivial Berry curvature. The longitudinal conductivity is modified from its standard expression because of Berry curvature effects. The $B$-dependent transverse conductivity also is modified by Berry curvature. Additionally, the transverse conductivity also comprises of a purely anomalous contribution even at zero $B$-field due to the Berry curvature. Thus the total contribution to the Nernst signal comprises of two parts: a $B$-dependent response, and a purely anomalous response. We derived analytic expressions for the Nernst coefficient in a linearized Dirac and Weyl Hamiltonian, and have also computed the total Nernst response for a lattice model of Weyl fermions numerically. We also pointed out that $B$-dependent normal Nernst signal is an even function of the chemical potential, but the anomalous Nernst coefficient is an odd function. As a result one would expect a slight asymmetry in the total Nernst response $\mu=0$, which is evident from our numerical studies. 

Additionally, we also examined the magneto-thermal conductivity of a WSM, and find that for orthogonal experimental setup, similar to Nernst experiment, the Wiedemann-Franz law holds for both longitudinal and Hall conductivities (normal and anomalous). For the parallel setup we find an additional $B^2$ dependence of the Lorenz number arising from the chiral anomaly term $\mathbf{v}\cdot\mathbf{\Omega}$. In a previous theoretical work~\cite{Kim2:2014}, both of these conclusions have been reported for a linearized WSM, and the violation of Wiedemann-Franz law in the parallel setup has been ascribed to the role of axion-electrodynamics because of the topological $\mathbf{E}\cdot\mathbf{B}$ term. We verify this violation of Wiedemann-Franz law in a lattice Hamiltonian, and it only depends on the presence of Berry curvature in a system, and therefore it is not an artifact of a linearized theory.

\textit{Acknowledgment:} G.S and S.T are supported by AFOSR (FA9550-13-1-0045). P.G was supported by NSF-JQI-PFC and
and LPS-CMTC.

\end{document}